\documentclass[manuscript,screen]{acmart}

\usepackage{pifont}
\usepackage{multirow}
\usepackage{array}
\usepackage{xspace}
\usepackage{placeins}
\usepackage{listings}
\usepackage{fontawesome5}   

\definecolor{simplexgreen}{HTML}{008000}
\definecolor{weightedorange}{HTML}{C65D00}

\newcommand{\onet}{O*NET\xspace}
\newcommand{\bench}{\citeauthor{eloundou2024gpts}\xspace}

\newcommand{\hficon}{\raisebox{-0.2em}{\includegraphics[height=1.1em]{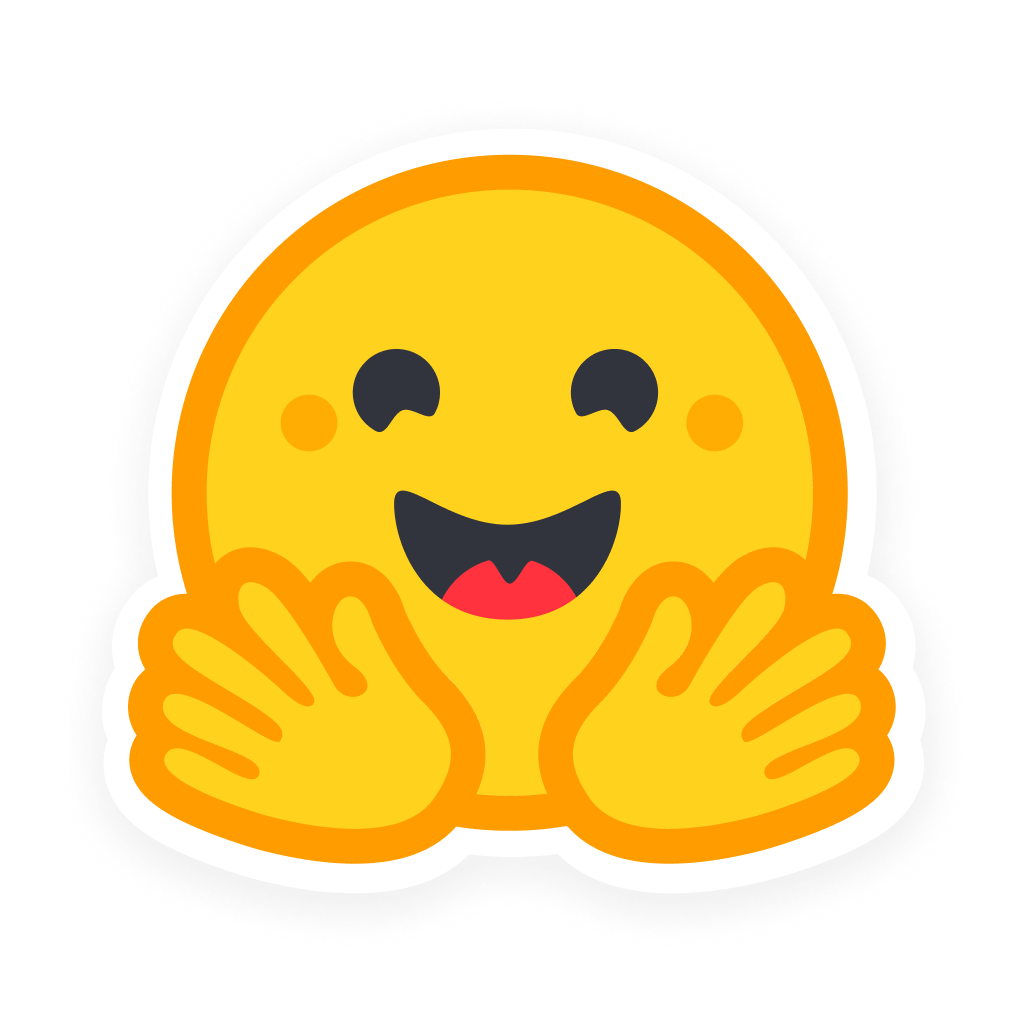}}}

\setcopyright{acmlicensed}
\copyrightyear{2027}
\acmYear{2027}
\acmDOI{XXXXXXX.XXXXXXX}
\acmConference[FAccT '27]{The 2027 ACM Conference on Fairness, Accountability, and Transparency}{June 21--24, 2027}{}
\acmISBN{978-1-4503-XXXX-X/2027/06}

\setcopyright{none}
\renewcommand\footnotetextcopyrightpermission[1]{}

\begin{document}

\title[Who Counts as Exposed?]{Who Counts as Exposed? Measurement Choices in an Open, Re-estimable Index of Occupational AI Exposure}

\author{Luca Mouchel}
\affiliation{%
  \institution{Massachusetts Institute of Technology}
  \city{Cambridge}
  \state{MA}
  \country{USA}}
\email{lucamo02@mit.edu}

\author{Pierre Bouquet}
\affiliation{%
  \institution{Massachusetts Institute of Technology}
  \city{Cambridge}
  \state{MA}
  \country{USA}}
\email{pibou149@mit.edu}

\author{Yossi Sheffi}
\affiliation{%
  \institution{Massachusetts Institute of Technology}
  \city{Cambridge}
  \state{MA}
  \country{USA}}
\email{sheffi@mit.edu}

\renewcommand{\shortauthors}{Mouchel et al.}

\begin{abstract}
Occupational AI exposure indices are used as inputs to economic analysis, workforce planning, and policy, and the choices built into them decide which workers are described as exposed. We build an open, re-estimable task-level index and vary its two main design decisions, context and calibration, to show how much the answer depends on them. For each of the 18,796 occupation--task pairs in \onet, seven open-weight reasoning models rate exposure under a common rubric. Each model rates with or without a context of retrieved news and research passages, workplace descriptors, and a bottleneck-based reasoning protocol, and the ensemble is combined with or without simplex-constrained calibration to observed Claude usage. The largest difference from the dominant 2023 benchmark is not a disagreement about what AI can do but about how much credit to give exposure that depends on software integrations not yet built, a choice that no data can settle. At the benchmark's own weight, the calibrated index counts fewer exposed workers than the benchmark, and when only direct exposure counts it counts more. At full score resolution, the ensemble ranks tasks better than the benchmark against observed usage and an independent physical-task classification. Adding context mainly shifts the level of exposure rather than improving the ranking. Context and calibration both relocate exposure toward higher-wage occupations, and calibration, which anchors the index to one platform's users, does more of it, so we also release the uncalibrated arm. Exposure indices used as policy inputs should state such choices, report their numbers under the alternatives, and be re-estimated as systems change.

\par\medskip\noindent
\faGlobe~\href{https://workanalyticslab.com}{\texttt{workanalyticslab.com}}\hfill
\faGithub~\href{https://github.com/MIT-Work-Analytics-Laboratory/RAG-Exposure}{\texttt{MIT-Work-Analytics-Laboratory/RAG-Exposure}}\hfill
\hficon~\href{https://huggingface.co/datasets/MIT-WAL/evidence-grounded-ai-exposure}{\texttt{Exposure index}}\,/\,\href{https://huggingface.co/datasets/MIT-WAL/ai-jobs-news-articles-abstracts}{\texttt{Corpus data}}
\end{abstract}



\maketitle

\section{Introduction}
\label{sec:introduction}

Generative AI can change the speed and quality of specific knowledge-work tasks \cite{noyzhang2023productivity,brynjolfsson2025generative}, and current systems can use browsers and software tools to carry out multi-step workflows under human direction. Because a technology may strongly affect one activity while leaving the rest of a job intact, research on AI and work has moved from the job level \cite{frey2017future,brandes2016opening} to the task level \cite{pioneer,webb2019impact,acemoglu2022artificial,eloundou2024gpts}, following the task-based view of technological change \cite{10.1162/003355303322552801}. The central question is then \emph{occupational exposure}: which tasks can be done much faster, at comparable quality, with help from an AI system. Exposure is not displacement, realized adoption, or a forecast of wages. It is a forward-looking signal of where an evolving technology is plausibly applicable.

Exposure estimates now inform international discussion of worker protection, reskilling, and the distributional consequences of AI. The OECD, the ILO, the IMF, and the World Economic Forum all rely on task-based exposure frameworks to describe which workers and sectors may be affected \cite{oecd2023employment,gmyrek2025genai,Cazzaniga2024Gen,WEF2025FutureJobs}. The dominant benchmark is \citet{eloundou2024gpts}, which classified \onet tasks into four exposure classes (E0--E3) using zero-shot judgments from GPT-4 in 2023. It established task-level exposure as a common empirical object. Yet its labels come from the internal prior of a single proprietary model, not from evidence about the task, the work setting, or demonstrated capability. The procedure did not require any inspectable reasoning, and the 2023 scores are still used in 2026 employment analysis \cite{brynjolfsson2025canaries,massenkoffmccrory2026labor}, in part because re-running the measurement was never part of the design. When a model-generated label is reused as a policy input, it becomes a measurement claim about the world, and the choices built into it decide whose work is described as at risk. Such a claim should be reproducible, externally grounded, inspectable, and renewed over time \cite{mouchel2026jobsaiexposuremeasured}, and its validity should be argued rather than assumed \cite{jacobs2021measurement}.

Existing measures fall into two families. \emph{Theoretical capability} measures ask whether an AI system could in principle help complete a task much faster \cite{eloundou2024gpts}. They identify where AI may become relevant before adoption is widespread, but they can overstate present exposure when they rely on model judgments that are not tied to evidence about current systems or working conditions. \emph{Observed usage} measures track which tasks are actually performed on a deployed platform, as in recent measures built from Claude and Bing interaction data \cite{massenkoffmccrory2026labor,tomlinson2025workingwithai}. They reflect real behavior, but low Claude usage does not mean that AI cannot help with a task done through another model, a specialized product, or an enterprise workflow that the platform cannot see.

This paper builds a third construct: a measure of present-day capability, grounded in a context a reader can inspect and released in a usage-free and a usage-anchored version. We use it to measure how much the answer to ``who is exposed'' depends on the choices inside the measure. For each of the 18,796 occupation--task pairs in \onet 30.2 \cite{onet_database_30_2}, seven open-weight reasoning models assess the task under a common rubric written for an up-to-date agentic system. Each model receives a \emph{context} with three parts: time-stamped news and research passages retrieved for the task, six \onet work-context descriptors for the occupation, and a protocol that asks it to decompose the task and label its time bottleneck. We calibrate the task scores against observed Claude usage with inspectable, simplex-constrained weights and aggregate them to occupations with survey-based task time shares \cite{measuringintensive}. A two-by-two design compares the \bench benchmark with every combination of context (with or without) and calibration (with or without). Because every output is stored, we recompute every headline number under the alternatives without new inference.

The headline result is about the measure, not about AI. Most of the gap between the ensemble's labels and the benchmark's lies in one class, exposure that would need software or integration that does not yet exist. How much credit that class receives is a normative choice with no empirical answer, and it decides whether the index counts more or fewer exposed workers than the benchmark. The ensemble ranks tasks with observed usage above tasks without it better than the benchmark does, although part of that advantage on all tasks comes from its finer score resolution. Context and calibration both relocate exposure toward legal, financial, and analytical work and toward higher wages, and in proportional terms calibration does more of this relocating than the context.

Our contributions are: (i) evidence that the choices inside an exposure measure, above all the credit given to integration-dependent exposure, change who is counted as exposed, and by how much, across occupations and wages; (ii) a traceable, multi-model framework that estimates task-level exposure from retrieved evidence, workplace descriptors, and an explicit reasoning protocol, released in a usage-free and a usage-anchored version; (iii) a two-by-two evaluation that separates what the context and the calibration each change, against observed usage and an independent physical-task classification; and (iv) a scheduled, incremental pipeline, so that any two vintages of the index can be compared task by task.

\section{Related Work}
\label{sec:related-work}

\paragraph{Task-based measurement of AI exposure.}
\citet{10.1162/003355303322552801} formalized the task-based view for computerization, showing that computers substitute for routine tasks while complementing nonroutine cognitive and interpersonal work. Early AI exposure measures mapped technical indicators of progress to occupational content, for instance by comparing the text of AI patents with task descriptions \cite{webb2019impact}. More recently, \citet{bouquet2026news} use sentiment from roughly two million news articles as a time-varying proxy for automation risk. The most influential task-level measure of LLM exposure is the E0--E3 framework of \citet{eloundou2024gpts}, which asks whether an LLM, alone or with complementary software, could halve task completion time without reducing quality. Its labels were produced by zero-shot prompting of a proprietary model, without retrieved evidence, workplace constraints, or reasoning instructions, so they carry no information about why a classification should remain valid as capabilities change.

\paragraph{Capability, observed usage, and adoption.}
\citet{tomlinson2025workingwithai} map Bing Copilot interactions to occupational activities, and \citet{massenkoffmccrory2026labor} use Claude interaction traces to construct task- and occupation-level measures of observed use. Usage data provide a behavioral check on theoretical exposure, since a technically suitable task may remain unused because of limited access, integration costs, or low demand, but interaction traces reflect one platform's users and deployment context.

\paragraph{Exposure indices as measurement instruments.}
An exposure score is a measurement of an unobservable construct, and, following \citet{jacobs2021measurement}, its validity has to be shown rather than assumed. Measurement theory adds that validity covers the consequences of using a score as well as its meaning \cite{messick1995validity}, and work on problem formulation shows that decisions about how a target is defined, often made for convenience, carry normative weight \cite{passi2019problem}. Benchmark datasets in machine learning have similar pitfalls, where unstated construction choices shape what a benchmark rewards \cite{blodgett2021salmon}. Exposure indices are now used to predict unemployment risk \cite{frank2025ai} and to direct policy attention, yet a systematic review finds that the proxies used for exposure are diverse and that different methods often reach contradictory conclusions \cite{ghosh2025aiexposure}. We treat the rubric, the credit given to integration-dependent exposure, the aggregation weights, and the calibration target as such construction choices, and report how sensitive the conclusions are to them.

\paragraph{Aggregation and grounding.}
\citet{eloundou2024gpts} weight tasks by whether they are core or supplemental, \citeauthor{massenkoffmccrory2026labor} ask Claude to estimate the time spent on each task, and \citet{measuringintensive} derive within-occupation task shares from \onet task-frequency responses. On the judgment side, \citet{mouchel2026jobsaiexposuremeasured} argue that exposure measures should be externally grounded, inspectable, and periodically revalidated, and describe a retrieval-augmented pipeline over the same \onet pairs. This paper adds the two-by-two design, the calibration, the sensitivity to the weight on integration-dependent exposure, and the wage incidence. Retrieval augmentation conditions model outputs on external evidence rather than on parametric memory alone \cite{lewis2021retrievalaugmentedgenerationknowledgeintensivenlp,shuster2021retrievalaugmentationreduceshallucination}, and explicit reasoning improves performance on complex judgments \cite{wei2022cot,kojima2022zeroshotcot}.

\section{Measurement Framework}
\label{sec:framework}

\paragraph{Tasks and the definition of exposure.}
\onet, sponsored by the U.S.\ Department of Labor, links detailed occupations to the tasks workers perform and surveys workers about how work is done \cite{onet_database_30_2}. We use the \onet task as the unit of measurement. We define AI exposure as \emph{present-day task applicability}: whether access to a contemporary AI system could reduce task completion time by at least half at equivalent quality, following \citet{eloundou2024gpts}. For comparability, tasks are assigned to four classes: (E0) no meaningful exposure, (E1) direct exposure through a general-purpose AI system, (E2) exposure requiring additional software or organizational integration, and (E3) exposure dependent on visual capabilities (Section~\ref{sec:taxonomy}). As in \bench, the classes are mapped to an exposure proxy $\beta$, the benchmark's own name for this quantity, not to a probability of automation:
\begin{equation}
\label{eq:betadef}
    \beta = \begin{cases}
        0 & \text{if E0}\\
        \delta & \text{if E2 or E3} \\
        1 & \text{if E1,}
    \end{cases}
    \qquad \delta=\tfrac{1}{2}.
\end{equation}
The weight $\delta$ on integration- and vision-dependent exposure is not an empirical quantity. It encodes how much credit a task receives for a speed-up that would need software or integration beyond the standalone system, and so how much unrealized potential counts as exposure today. We follow \bench in setting $\delta=\tfrac{1}{2}$ for comparability, and Section~\ref{sec:delta} reports every headline number for $\delta$ from 0 to 1. Exposure is distinct from adoption and from labor-market outcomes. A task may be technically exposed even if AI is not yet deployed in the workplace.

\paragraph{From task to occupation.}
\label{sec:tasktime}
A simple average of task scores would give the same importance to tasks performed continuously and occasionally. We instead use the task-share estimates of \citet{measuringintensive}, which convert \onet survey responses on task relevance and frequency into normalized within-occupation shares $\pi_{k,i}$. These shares approximate how work time is allocated, with $\sum_{i\in\mathcal{T}_k}\pi_{k,i}=1$ over the task set $\mathcal{T}_k$ of occupation $k$. Occupational exposure is the weighted sum
\begin{equation}
{\varepsilon}_{k}
=
\sum_{i \in \mathcal{T}_k}\pi_{k,i}{\beta}_{k,i}.
\label{eq:occupation-exposure}
\end{equation}
This is an index of task-time-weighted applicability that can be traced back to its components. It is not the share of an occupation that will be automated.

\paragraph{Grounding, calibration, and the gap between the two families.}
An ungrounded language-model response relies largely on the model's prior associations between a task and AI. We instead assess each task in a context made of retrieved evidence about current AI capabilities and deployments, its \onet workplace descriptors, and a reasoning protocol that ties the label to the task's time bottleneck. Several open-weight reasoning models then evaluate the same prompt. External evidence can reduce reliance on parametric knowledge, the descriptors and the protocol make the basis of the label explicit, the ensemble reduces dependence on any single model's prior, and open weights make the measurement reproducible. Because the grid of Equation~\ref{eq:betadef} is coarse, we then calibrate the model assessments at the task level against observed usage from the Anthropic Economic Index \cite{massenkoffmccrory2026labor}. Usage is a useful behavioral signal, because technical capability does not automatically turn into workplace adoption. But it is an anchor rather than ground truth, since low usage may reflect limited applicability, slow adoption, platform coverage, or use elsewhere. Anchoring pulls the weights toward the models whose scores co-vary with adoption, but with simplex weights and no intercept it can only mix the models' own levels, and on a target that is mostly zero it mixes toward the most conservative model (Section~\ref{sec:calibration}). We therefore also release the uncalibrated ensemble as an index in its own right. The measure is meant to sit between potentially overstated theoretical capability, such as the 48\% exposure that \bench assign to general pediatricians, and potentially understated platform-specific use, as for legal professionals who rely on domain-specific services.

\section{Data}
\label{sec:data}

\paragraph{\onet occupations, tasks, shares, and context.}
Version 30.2 of \onet covers 923 occupations and 18,796 occupation--task pairs, and links each task to broader Intermediate and Detailed Work Activities (IWAs and DWAs). A task that reads as suitable for an AI system may in practice depend on physical access, real-time interaction, exact execution, or responsibility for consequential outcomes. We therefore attach six occupation-level descriptors from \onet work-context surveys to every task in an occupation: the frequency of face-to-face discussion, responsibility for the health and safety of others, the degree to which the job is automated, the importance of being exact or accurate, the importance of continuous repetitive physical or mental activity, and the frequency of written letters and memos. Each is on a 0--100 scale with its response category (Appendix~\ref{app:prompt-format}). These descriptors characterize how the work is normally done, not exposure.

\paragraph{External retrieval corpus.}
The news collection was assembled from GDELT 2.0, a global news-monitoring database that records structured metadata and a link for each report \cite{leetaru2013gdelt}. Monthly BigQuery queries selected English-language records from a curated allowlist of 47 major news, business, technology, and science outlets, keeping only records whose themes or named entities indicate a link to AI, automation, or emerging technology (Appendix~\ref{app:gdelt-query}). We retrieved the article text with an open-source scraper. The collection contains more than 32,000 articles published between January 2025 and June 2026. The academic collection was gathered through the Semantic Scholar bulk-search interface \cite{Kinney2023TheSS} with four query families on the intersection of AI, employment, labor markets, and occupations. It is restricted to journal articles and reviews in economics, business, sociology, and computer science from 2020 onward (Appendix~\ref{app:s2-query}), and contains roughly 21,000 abstracts, which gives 53,526 documents in total. We treat neither source as proof that a system can perform a task, since reporting may be promotional and is likely to be uneven across occupations (Section~\ref{sec:threats}).

\paragraph{Theoretical-exposure benchmark.}
The benchmark is the task-level exposure dataset of \citet{eloundou2024gpts}, which contains GPT-4 task judgments and occupation-level aggregates under the E0--E3 taxonomy and the mapping of Equation~\ref{eq:betadef}. Because our measure uses the same taxonomy, its outputs can be compared with the benchmark task by task. The benchmark is also used to build Anthropic's usage measure, so the two are not independent.

\paragraph{Anthropic Economic Index.}
\citet{massenkoffmccrory2026labor} map anonymized Claude interactions to \onet tasks, combining work-related Claude.ai conversations with first-party API traffic. A task must have at least 100 observations, about 0.0025 percent of traffic, to pass a privacy and reliability gate. The task-level quantity that Anthropic calls \emph{penetration} is
\begin{equation}
\widetilde{r}_{i}
=
\mathbf{1}\!\left\{\mathrm{WorkUsage}_{i}\geq100\right\}
\times
\mathbf{1}\!\left\{\beta^{\mathrm{E}}_{i}\geq0.5\right\}
\times
\alpha_{i},
\label{eq:anthropic-task-penetration}
\end{equation}
where $\beta^{\mathrm{E}}_i$ is the \bench task score and $\alpha_i\in[0.5,1]$ is an automation factor that weights automative use above augmentative use (Appendix~\ref{app:anthropic}). The second indicator is a feasibility filter taken from the benchmark. Observed use alone cannot produce positive penetration for a task that the benchmark rated E0, so the behavioral target partly depends on the theoretical measure it is used to evaluate. Anthropic aggregates penetration to occupations as $R_k = \sum_{i\in\mathcal{T}(k)} w_i\widetilde{r}_i / \sum_{i\in\mathcal{T}(k)} w_i$, with $w_i$ an estimated task-time fraction. The task-level $\widetilde r_i$ is the calibration target for the ensemble. The occupation-level $R_k$ is not used to fit anything and is kept for a separate comparison after aggregation. Because $R_k$ aggregates the same task-level penetration that the calibration sees, Section~\ref{sec:occupation-alignment} also reports a version in which the weights for each occupation are fitted without any of its tasks.

\section{Method}
\label{sec:method}

The measure is a retrieval-augmented pipeline (Figure~\ref{fig:pipeline} in Appendix~\ref{app:config}). For every occupation--task pair, we give an ensemble of reasoning models the most relevant passages from the evidence corpus, together with the task and its workplace context. We then combine the model scores with calibrated weights and aggregate the calibrated task scores to occupations.

\subsection{Indexing and task-conditioned hybrid retrieval}

The unit of retrieval is a passage. Abstracts and short articles are kept whole, and longer articles are split semantically \cite{kamradt2024semantic,chroma2024evaluating_chunking} into chunks of at least 800 characters, which gives an index of over 176,000 passages encoded with BGE-M3 \cite{chen2024m3}. The query for a pair concatenates the occupation title, its description, the IWA and DWA, and the task statement, so that retrieval is conditioned on the occupation and not on the task string alone. We retrieve 40 candidates by dense similarity, rescore them with a hybrid of dense and sparse lexical similarity \cite{bruch2023analysis}, rerank them with a cross-encoder \cite{li2023making,nogueira2019passage}, and keep up to five above a relevance threshold (Appendix~\ref{app:config}). Receiving no passage is a valid outcome rather than an error.

\subsection{Grounded task inference}
\label{sec:taxonomy}

Each model receives the occupation title and description, the task statement, the IWA and DWA, the retrieved passages, and the panel of six \onet descriptors. The model is told to treat the descriptors as occupation-level priors that must not override the activity described by the task. All models apply the same 2026 rubric, which updates the E-label definitions of \bench for current frontier systems. The counterfactual system is a frontier agentic chatbot with general reasoning, planning, and public-web browsing, which can also work in standard browser-based and office-productivity environments through the worker's authorized access. It has no custom enterprise integrations, proprietary data, physical sensors or actuators, or organization-specific back-end access. E0 means that the threshold cannot be reached, which includes tasks dominated by physical execution or in-person requirements. E1 means that the standalone system reaches it. E2 means that it could be reached only with additional AI-enabled software, deeper enterprise integration, or internal data access. E3 means that it is reached only when integrated image or video understanding removes a binding visual constraint. The full system prompt, which is the measurement instrument, is in Appendix~\ref{app:system-prompt}. Each model must split the task into two to five concrete subtasks, assign a mode to each, and set the overall label from the bottleneck that dominates the task's time. In this way, a minor component that AI can speed up cannot decide the label of a task whose main bottleneck is not exposed. We use grammar-constrained decoding \cite{dong2025xgrammar} to force every response into a JSON schema with four fields: a one-paragraph rationale, a binary indicator of whether the retrieved context was relevant, the source numbers used, and the exposure label (Appendix~\ref{app:schema}). The output of model $m$ for pair $(k,i)$ is mapped to $\beta^{(m)}_{k,i}$ by Equation~\ref{eq:betadef}. Inference uses $M=7$ open-weight reasoning models from six developers (Section~\ref{sec:setup}).

\subsection{Behavioral calibration and aggregation}
\label{sec:calibration}

Calibration learns how to combine the model-specific scores so that their weighted average aligns with Anthropic's observed task penetration $y_i=\widetilde{r}_i$. Anthropic reports penetration by task text without the occupation, so when the same task appears in several occupations, we average its score within each model (2.26\% of unique tasks).

A positive penetration score is direct evidence that Claude has been used for a task. A zero may mean that the task is genuinely not performed with AI, or that adoption is slow, specific to other platforms, or too rare to appear. Only 7.6\% of matched tasks have positive penetration, so treating every zero as an equally reliable negative would make a capability-oriented measure reproduce the blind spots of a single usage source. We use the ensemble models themselves to tell the two kinds of zero apart. Tasks with positive penetration receive full weight. For zeros, the weight decreases with the mean exposure assigned by the models:
\begin{equation}
\omega_i
=
\begin{cases}
1 & \text{if }y_i>0,\\[3pt]
1-\dfrac{1}{M}\displaystyle\sum_{m=1}^{M}\beta_i^{(m)} &\text{if } y_i=0.
\end{cases}
\label{eq:linear-zero-weight}
\end{equation}
This choice introduces a deliberate circularity. The reliability given to each zero depends on the ensemble's own mean prediction, so the calibration objective partly defers to the judgments it is meant to correct. Appendix~\ref{app:sensitivity} refits the weights with every zero at full weight. Because 92\% of the target is zero, either version of the objective favors models that rate conservatively, which is one reason we report the equally weighted ensemble alongside the calibrated one.

The matched sample is split 80/20, stratified on positive penetration, and the weights are estimated only on the training set:
\begin{equation}
\mathbf{w}^{\star}
=
\operatorname*{arg\,min}_{w_1 \hdots w_M}
\left\{
\sum_{i\in\mathcal{I}_{\mathrm{train}}}
{\color{weightedorange}\omega_i}
\left(
y_i-\sum_{m=1}^{M}w_m\beta_i^{(m)}
\right)^2
\;\middle|\;
{\color{simplexgreen}
w_m\geq0\:\:\forall m,
\quad
\sum_{m=1}^{M}w_m=1
}
\right\}.
\label{eq:constrained-calibration}
\end{equation}
No intercept is included, so a task that every model rates as unexposed receives a calibrated score of zero. A negative weight would mean that a higher exposure rating from a model lowers the final score, which has no sensible interpretation. The constraint gives such a model zero weight instead. The sum-to-one constraint makes the calibrated score $\beta^\star_i=\sum_m w^\star_m\beta_i^{(m)}$ a weighted average that stays in $[0,1]$, and lets the reader see each retained model's contribution directly. Two consequences follow. Because the weights form a convex combination with no intercept, the calibrated score is a mixture of the models' existing levels, so its mean must lie between the lowest and the highest model mean, 0.205 and 0.369 on the calibration sample. The target's mean is 0.067, so the fit cannot match the level of usage and can only choose which levels to mix. And on a target that is 92\% zero the squared error is smallest for the mixture closest to zero, so the fit selects the most conservative model (Section~\ref{sec:weights}). The zero-weighting of Equation~\ref{eq:linear-zero-weight} limits this pull but does not remove it. We report the uncalibrated mean, $w_m=1/M$, alongside it. The calibrated score of each normalized task is assigned back to every pair that contains it and aggregated with the task shares as in Equation~\ref{eq:occupation-exposure}. An occupation score can therefore be broken down into its task shares, calibrated task scores, individual model labels, retrieved passages, and occupational context.

\subsection{Keeping the measure current}
\label{sec:recursive}

The corpus covers a time window, the rubric describes the systems available when it was written, and the calibration target is one platform's usage over one period, so every number above holds at one point in time. An index that is expensive to recompute is treated as a constant whatever its authors intend, so the pipeline is a scheduled job that writes a new dated release and keeps previous ones for task-by-task comparison. Each stage is incremental: only new documents are indexed, a pair is re-evaluated only if its top five passages change, and the ensemble weights are then refitted.

\section{Experimental Setup}
\label{sec:setup}

\paragraph{Model pool and decoding.}
Inference uses seven open-weight reasoning models in the 14 to 32B parameter range from six developers: DeepSeek-R1-Distill-Qwen-32B \cite{deepseek2025r1}, Gemma~4 26B-A4B \cite{gemmateam2026gemma4}, Ministral~3 14B \cite{liu2026ministral}, Qwen3 30B-A3B \cite{yang2025qwen3}, GPT-OSS-20B \cite{openai2025gptoss120bgptoss20bmodel}, Nemotron-3-Nano-Omni 30B-A3B \cite{nvidia2026nemotron3nanoomni}, and OpenReasoning-Nemotron 32B \cite{ahmad2025opencodereasoningiisimpletesttime,toshniwal2025genselect,moshkov2025aimo2winningsolutionbuilding} (Table~\ref{tab:model-pool} in Appendix~\ref{app:config}). We chose them so that the measurement can be reproduced, so that the ensemble spans six developers and different post-training regimes, and so that each model fits on a single GPU. Three of the seven share Qwen lineage, since the DeepSeek distillation and OpenReasoning-Nemotron are built on Qwen bases. All models run in their reasoning configuration at temperature 0.6, nucleus probability 0.9, and top-$k$ of 30. We draw one sample per model per task and serve each model with vLLM \cite{kwon2023efficient} on one NVIDIA H100 80GB GPU, so we do not measure within-model sampling variance. The two inference runs, with and without context, total roughly 263,000 generations.

\paragraph{Experimental arms.}
Five specifications appear throughout the results (Table \ref{tab:ablation-summary}). The benchmark $B_0$ is the published GPT-4 task rating of \citet{eloundou2024gpts}. The ensemble arms fully cross context with calibration: $A_0^{u}$ (no context, equal weights), $A_0$ (no context, calibrated), $A_1$ (context, equal weights), and $A_2$ (context, calibrated). We release both context arms: $A_1$ is the usage-free index and $A_2$ its usage-anchored variant, and the headline numbers are reported for both. The no-context arms remove the whole context block: the retrieved passages, the \onet descriptor panel, and the instruction block that requires a subtask decomposition. This block is replaced by a shorter one that asks for the label and a one-paragraph rationale, as in the context arm, without the evidence and descriptor rules (Appendix~\ref{app:instructions}). They keep the occupation fields, the rubric, the models, and the decoding. The context contrast therefore tests the passages, the descriptors, and the protocol together and says nothing about retrieval specifically. The two uncalibrated arms, $A_0^{u}$ and $A_1$, take the unweighted mean of the seven scores and are never fitted to Anthropic data, which makes their contrast the cleanest test of the context and the key comparison for any circularity objection. Every comparison with $B_0$ changes more than one thing at once. All four ensemble arms use the 2026 rubric and a different set of models, while $B_0$ is GPT-4 under the 2023 rubric. Differences between an ensemble arm and $B_0$ therefore measure the combined effect of rubric, models, and prompt, and only contrasts among the four ensemble arms isolate the context or the calibration.

\paragraph{Calibration and evaluation protocols.}
All arms are compared on one common held-out fold: the 3,391 test tasks of the with-context sample that also have a no-context and a benchmark score, of which 257 have positive penetration. For task-level comparisons, the no-context weights are fitted on the same training fold, so no arm sees a test task during fitting (Appendix~\ref{app:splits}). We solve Equation~\ref{eq:constrained-calibration} with SLSQP \cite{kraft1988slsqp}. Differences in AUC, average precision, and correlation between arms use 5,000 paired bootstrap resamples, with percentile 95\% intervals and two-sided $p$-values. Prometheus~2 (7B), an open evaluator trained for rubric-anchored assessment \cite{kim2024prometheus}, scores rationales from the context arm for groundedness and faithfulness on a five-point Likert scale \cite{likert1932technique}. Pairwise adjudication exercises run during development are summarized, with their limitations, in Appendix~\ref{app:adjudication}, but we do not use them as evidence.

\section{Results}
\label{sec:results}

\subsection{The resulting index}
\label{sec:the-index}
\label{sec:weights}
\label{sec:label-distribution}

\paragraph{Pipeline diagnostics.}
\label{sec:pipeline-diagnostics}
\label{sec:coverage}
Retrieval rarely fails. Of the 18,764 pairs for which a bundle was built, 99.6\% received at least one passage and 79.3\% received the full five. Rationales score $4.19 \pm 0.61$ out of 5 for \emph{groundedness} and $3.98 \pm 0.58$ for \emph{faithfulness} to the evidence bundle, which says nothing about whether the corpus is correct.

\paragraph{Calibration weights.}
The simplex constraint makes the fitted weights easy to read. Qwen3 30B-A3B receives $0.511$, OpenReasoning-Nemotron 32B $0.164$, Gemma~4 26B-A4B $0.160$, Ministral~3 14B $0.084$, and DeepSeek-R1-Distill-Qwen-32B $0.081$. GPT-OSS-20B and Nemotron-3-Nano-Omni receive zero. The final index is mostly a Qwen3 index, with four other models acting as a correction. Qwen3 is also the model with the lowest level, a mean of 0.205 on the calibration sample against 0.288 to 0.369 for the other six, and no mixture of these levels can reach the target's mean of 0.067 (Section~\ref{sec:calibration}), so the calibration selects the most conservative model. The calibrated mean of 0.250 sits between Qwen3 and the rest. Bootstrap resamples put a 95\% interval of $[0.485, 0.540]$ on the Qwen3 weight, but its size depends on how many positive tasks the fit sees (Appendix~\ref{app:weight-stability}).

\paragraph{Label distribution.}
The grounded labels (Figure~\ref{fig:exposure_dist} in Appendix~\ref{app:flows}) do not differ from the benchmark in the way one would expect from a uniformly more conservative measure. On direct exposure, E1, six of the seven grounded models exceed the benchmark's 13.5\%. Almost the entire drop in overall level comes from E2/E3, the claim that the threshold \emph{could} be reached with additional software, deeper integration, or visual capability. The benchmark applies this class to 41.4\% of tasks, and the grounded models to only 12\% to 24\%. What the measure is much more reluctant to do is to count a hypothetical future integration as present exposure. Appendix~\ref{app:flows} reports the full label flow with examples.

\subsection{Alignment with observed Claude usage}
\label{sec:behavioral}

Recall that penetration is gated on the benchmark's own feasibility filter (Equation~\ref{eq:anthropic-task-penetration}). No task that the benchmark rated E0 can have positive penetration, so the benchmark never produces a false negative, and any task that we rate E0 but that has positive usage counts against us. Table~\ref{tab:ablation-summary} collects the results for all five arms, and Appendix~\ref{app:detailed-tables} gives confidence intervals and the contrasts between arms.

\begin{table}[t]
\centering
\small
\setlength{\tabcolsep}{3.5pt}
\begin{tabular}{llccccc}
\toprule
& & $B_0$ & $A_0^{u}$ & $A_0$ & $A_1$ & $A_2$ \\
& Context & -- & \ding{55} & \ding{55} & \ding{51} & \ding{51} \\
& Calibrated & -- & \ding{55} & \ding{51} & \ding{55} & \ding{51} \\
\midrule
\multirow{3}{*}{\shortstack[l]{Task\\(held out)}}
& AUC, all tasks ($n=3{,}391$) & 0.766 & 0.813$^{***}$ & \textbf{0.821}$^{***}$ & 0.797$^{**}$ & 0.791 \\
& AUC, feasible tasks ($n=1{,}863$) & 0.543 & 0.653$^{***}$ & \textbf{0.670}$^{***}$ & 0.642$^{***}$ & 0.642$^{***}$ \\
& Average precision & 0.152 & 0.204$^{***}$ & 0.215$^{***}$ & \textbf{0.223}$^{***}$ & 0.221$^{***}$ \\
\midrule
\multirow{2}{*}{\shortstack[l]{Occupation\\($n=756$)}}
& Pearson $r$ with $R_k$ & 0.640 & 0.602$^{*}$ & 0.635 & 0.662 & \textbf{0.686}$^{*}$ \\
& Spearman $\rho$ with $R_k$ & 0.701 & 0.690 & 0.688 & 0.700 & \textbf{0.702} \\
\midrule
Independent target & AUC, non-physical vs.\ physical & 0.815 & \textbf{0.915}$^{***}$ & 0.914$^{***}$ & 0.885$^{***}$ & 0.866$^{***}$ \\
\midrule
Level & Mean task score ($n=18{,}200$) & 0.348 & 0.465 & 0.377 & 0.311 & 0.257 \\
\midrule
& Fitted to Anthropic data? & no & no & yes & no & yes \\
\bottomrule
\end{tabular}
\caption{The five arms across all outcomes. Task-level rows use the common held-out fold, with 257 positives. \emph{Feasible tasks} are those the benchmark rates $\beta^{\mathrm{E}}\geq0.5$, which include every positive. The independent target is the physical-task classification of \citet{tomlinson2025workingwithai} (18,183 pairs). The mean task score is over the 18,200 occupation--task pairs that have a benchmark label and a score under every arm; it is descriptive and has no best value. Bold marks the best value in each row. Major-group results are in Appendix~\ref{app:detailed-tables}. Stars mark significance against $B_0$ from paired bootstraps, with $^{*}p<0.05$, $^{**}p<0.01$, $^{***}p<0.001$. A star does not mean that the difference favors the arm: the occupation-level $r$ of $A_0^{u}$ is significantly \emph{lower}.}
\label{tab:ablation-summary}
\end{table}

\paragraph{Task-level discrimination.}
\label{sec:task_auc}
Because only 7.6\% of matched tasks have nonzero penetration, we ask whether a score ranks tasks with observed usage above tasks without it. Every ensemble arm has a higher AUC than the benchmark, by $+0.025$ to $+0.055$, significantly so for all arms but $A_2$ ($p=0.055$), and a higher average precision, by $+0.053$ to $+0.071$ (all $p<0.001$). The gate hides how large this advantage is. Every positive task is one that the benchmark already rated feasible, and within the 1,863 feasible held-out tasks the benchmark is close to chance (AUC 0.543), while every ensemble arm reaches 0.642 to 0.670 (all $p<0.001$). This ranking advantage comes from the ensemble, not from the context. Adding context changes the AUC by $-0.015$ without calibration and by $-0.030$ with it, and the average precision by $+0.018$ and $+0.006$. Only the drop in AUC for the calibrated pair is significant. Part of the advantage on all tasks is score resolution: $B_0$ takes three values and cannot order tasks within a label, while the ensemble scores take many. Rounded to the same grid, the ensemble arms' all-task AUCs fall to between 0.748 and 0.788 and are no longer distinguishable from the benchmark's 0.766, but their feasible-set AUCs stay at 0.598 to 0.639 against 0.543 (all $p<0.01$), so what the ensemble adds is separation among tasks the benchmark already rates feasible, not finer ties alone (Appendix~\ref{app:detailed-tables}).

\paragraph{Occupation and industry level.}
\label{sec:occupation-alignment}
Aggregating the task scores with the \onet time shares and comparing them with Anthropic's occupation-level $R_k$ is a stricter test. $R_k$ is not used to fit weights or to select any configuration, and the two uncalibrated arms involve no fitting at all. Here the context matters more than calibration. Adding context raises the Pearson correlation by $+0.059$ without calibration and by $+0.051$ with it, while calibration adds $+0.032$ without context and $+0.024$ with it (all $p<0.001$). $A_2$ has the strongest association with observed usage ($r=0.686$ against $0.640$ for the benchmark, $\Delta r=+0.046$, $95\%$ CI $[0.009, 0.081]$, $p=0.013$), a modest margin whose interval barely excludes zero, and the same ordering holds across the 22 major groups (Appendix~\ref{app:detailed-tables}). Because $R_k$ aggregates the task-level target, we also fit the weights for each fifth of occupations without any of their tasks. The cross-fitted index has the same correlation ($r=0.686$, $\Delta r=+0.045$, 95\% CI $[0.010, 0.080]$), so the result is not an artifact of leakage (Appendix~\ref{app:sensitivity}). The Spearman correlations stay between 0.688 and 0.702 for all arms. The gain is therefore not in the rank order of occupations but in the \emph{spacing} of the scores, which is what matters for policy uses that rely on the values rather than the ranks.

\paragraph{Ranking versus level.}
\label{sec:ablation-synthesis}
The arms separate two properties of the measure. How well it \emph{ranks} tasks is a property of the ensemble, which beats the benchmark with or without context at full score resolution, here and in Section~\ref{sec:triangulation}. Where it sets the \emph{level} depends on the context. Without context, the mean task score is 0.465, a third higher than the benchmark's 0.348. With context it is 0.311, and calibration lowers it to 0.257, because a fit to a target that is 92\% zero rewards the most conservative model (Section~\ref{sec:weights}). No arm wins every row of Table~\ref{tab:ablation-summary}, and $A_2$ wins only the two occupation-level rows: the Spearman correlation by a trivial 0.001, and the Pearson correlation by $\Delta r=+0.046$ with an interval that barely clears zero. Its task-level AUC is not significantly above the benchmark's ($p=0.055$), and both no-context arms rank tasks and separate physical work better. We therefore do not single out one arm as the index: $A_2$ serves users who want the values, not only the ranks, aligned with one platform's adoption. The anchor buys that modest gain at a cost. The fit puts half of the weight on the most conservative model (Section~\ref{sec:weights}), that weight drifts from 0.50 to 0.63 with the number of positive tasks and collapses under class balancing (Appendix~\ref{app:weight-stability}), and the anchor, more than the context in proportional terms, moves exposure toward the occupations in which the platform's users work (Section~\ref{sec:what-changes}). Which arm to use is itself one of the measurement choices of Section~\ref{sec:discussion}.

\subsection{Evidence that does not come from Anthropic}
\label{sec:triangulation}
\label{sec:physical}
\label{sec:winrates}

A critic could still argue that the measure has been shaped to resemble Claude usage through the rubric, the corpus, and the choice of what counts as evidence; a test in which Anthropic data plays no part addresses this. The clearest external boundary on present AI capability is physical execution. \citet{tomlinson2025workingwithai}, working from Bing Copilot usage, classify \onet tasks as physical or non-physical with a GPT-5 classifier validated against human annotators. Every ensemble arm separates the two classes better than the benchmark, with AUCs of 0.866 to 0.915 against 0.815 (all $p<0.001$). Rounded to the benchmark's grid, $A_0^{u}$, $A_0$, and $A_1$ still separate the classes better than the benchmark (0.890, 0.842, and 0.827), but $A_2$ falls to 0.767, below it, because calibration also pushes many non-physical tasks under the E0 cut-off (Appendix~\ref{app:independent}). The context again acts on the level more than on the ranking. It lowers the AUC by 0.031 to 0.048, but it raises the share of physical tasks scored at 0.10 or below from 68.5\% to 83.5\% without calibration and from 87.4\% to 89.8\% with it, against 79.3\% for the benchmark (Appendix~\ref{app:independent}). Pairwise adjudications run during development are summarized in Appendix~\ref{app:adjudication}. Because their item sets were curated, we do not treat them as evidence.

\subsection{What the measure changes}
\label{sec:what-changes}
\label{sec:exposure_relocation}
\label{sec:overview-inbetween}

\begin{figure}[t]
    \centering
    \begin{minipage}[c]{0.495\linewidth}
        \centering
        \includegraphics[width=\linewidth]{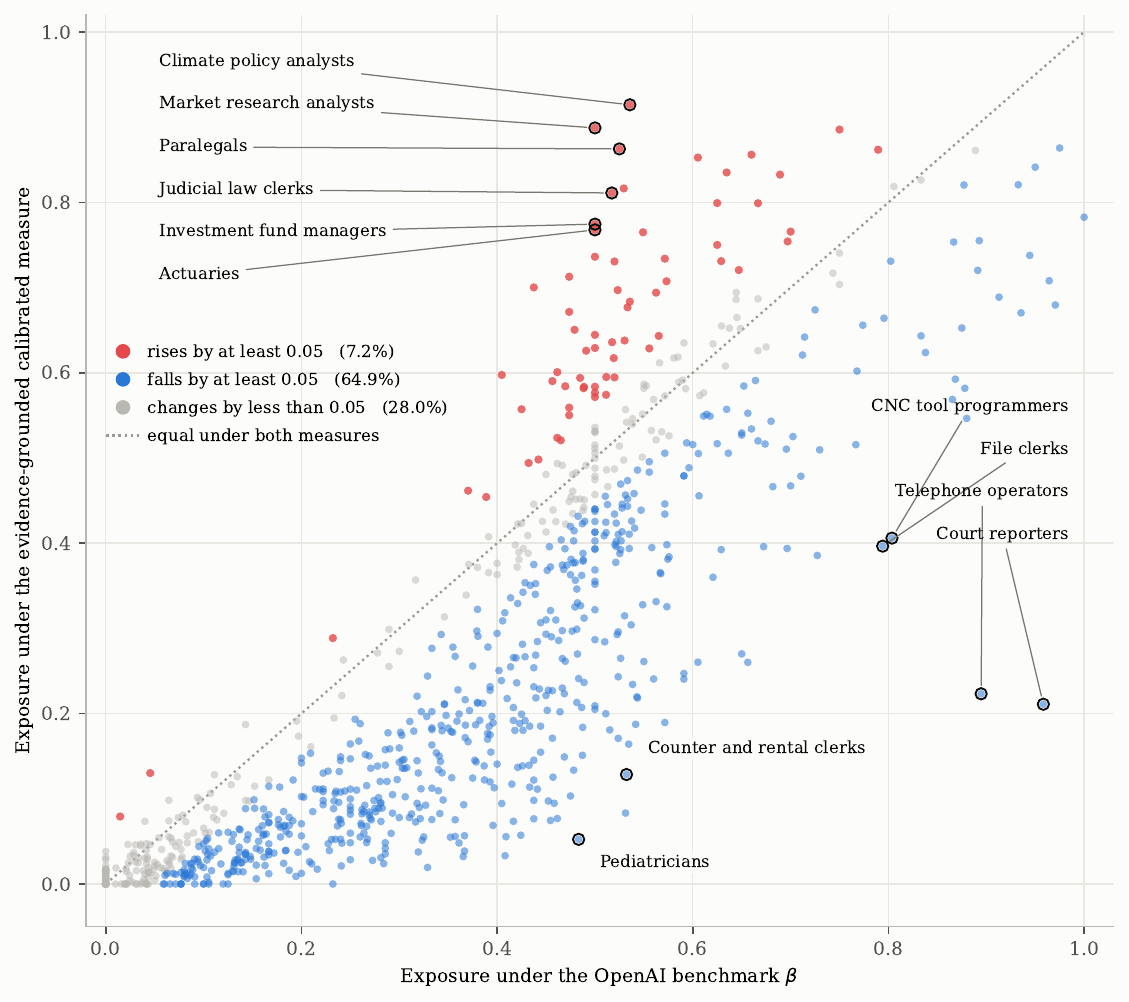}
    \end{minipage}\hfill
    \begin{minipage}[c]{0.495\linewidth}
        \centering
        \includegraphics[width=\linewidth]{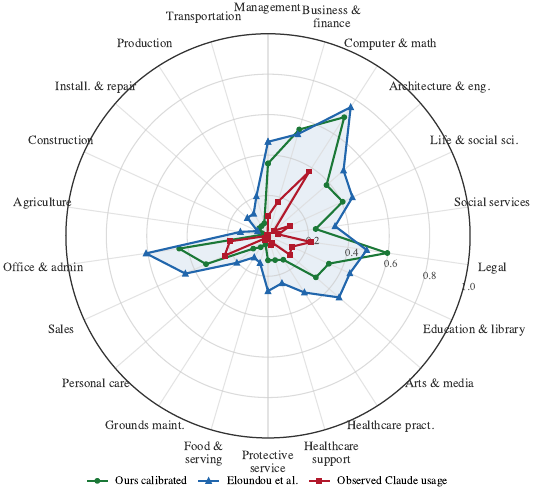}
    \end{minipage}
    \caption{\emph{Left:} detailed-occupation exposure under the benchmark $\beta$ and the calibrated grounded measure. Red points rise by at least 0.05, blue points fall by at least 0.05, and grey points change by less. Labeled occupations are the largest movers. \emph{Right:} major-group exposure by source (equal-weighted means over six-digit SOC occupations): the \bench benchmark (blue), the calibrated index $A_2$ (green), and observed Claude usage (red). The index lies below the benchmark in 20 of the 22 groups and above it in legal and in business and financial operations.}
    \Description{Two panels. The left panel is a scatter plot of 923 occupations, with benchmark exposure on the horizontal axis and calibrated exposure on the vertical axis. Most points fall below the diagonal. Labeled points that rise include market research analysts, paralegals, judicial law clerks, and actuaries. Labeled points that fall include court reporters, telephone operators, file clerks, and pediatricians. The right panel is a radar chart over 22 major occupational groups with three closed lines. The observed Claude usage line is the innermost in every group. The benchmark line is the outermost in 20 groups, and the calibrated index crosses outside it in the legal and the business and financial groups.}
    \label{fig:relocation-radar}
\end{figure}

The calibrated index does not simply rescale the benchmark downward. Across 923 occupations, mean exposure falls from $0.345$ under the benchmark to $0.297$ under $A_1$ and $0.245$ under $A_2$, but the benchmark and the calibrated score remain strongly related (Pearson $0.853$, Spearman $0.911$). The revisions are structured (Figure~\ref{fig:relocation-radar}, left), with 64.9\% of occupations falling and 7.2\% rising by at least 0.05. The largest downward revisions are physically constrained or built around real-time interaction, with court reporters falling from $0.96$ to $0.21$ and general pediatricians from $0.48$ to $0.05$. The largest upward revisions are analytical and document-intensive, with market research analysts rising from $0.50$ to $0.89$ and paralegals from $0.53$ to $0.86$. Most label changes in both directions (71.5\%, Appendix~\ref{app:flows}) start in the benchmark's large tool-dependent E2 class. The direction of a move predicts observed usage within each benchmark class (Appendix~\ref{app:flows}), but the measure also produces false negatives: 12.5\% of the benchmark-E2 tasks that the calibrated score moves to E0, and 8.6\% of the E1 tasks, still show Claude usage, and 304 of the 1,288 calibration tasks with positive penetration (23.6\%) are rated E0 by $A_2$, 159 (12.3\%) by $A_1$, with all seven models unanimous in only 46 cases. Most of these tasks have a drafting or documentation component inside work whose bottleneck is elsewhere, and observed use does not require that the task's time be halved, so a low score should be read as the absence of a time-halving path under the rubric, not of any use (Appendix~\ref{app:flows}).

Weighted by national employment \cite{bls_oews}, the calibrated index places 30.1 million U.S.\ workers (23.0\%) in occupations with exposure above $0.5$, against 36.2 million (27.7\%) under the benchmark. The context moves this number in one direction only. The calibrated no-context arm places 49.9 million workers (38.2\%) above $0.5$, and adding context moves 136 occupations, with 19.8 million workers, below the threshold and none above it. Without calibration, context lowers the count from 69.0 million (52.8\%) to 39.0 million (29.9\%). Figure~\ref{fig:relocation-radar} (right) compares the calibrated index with the benchmark's $\beta$ and with observed usage across the 22 major groups. The index lies below the benchmark in 20 groups and above it in two, legal (0.60 against 0.49) and business and financial operations (0.55 against 0.52); the largest gaps below the benchmark are in healthcare practitioners, office and administrative support, and protective service. That it lies above observed usage in every group says little, since usage on one platform is a lower bound for applicability by construction.

The correction is not uniform across the wage distribution. With median wages from the same OEWS release, mean exposure in the lowest-paid fifth of workers falls from 0.177 under the benchmark to 0.127 under $A_1$ and 0.086 under $A_2$, while in the highest-paid fifth it moves from 0.498 to 0.493 and 0.418 (Appendix~\ref{app:wages}). The absolute gap between the two fifths therefore barely moves, at 0.32, 0.37, and 0.33, and the steeper gradient is proportional: the bottom is cut by half under $A_2$ and by a quarter under $A_1$, the top by a sixth and by almost nothing. The ratio of top-quintile to bottom-quintile exposure rises from 2.8 under the benchmark to 4.9, and in proportional terms calibration does more of this than the context: the ratio is 2.9 for $A_0^{u}$, 3.9 for $A_1$, 4.7 for $A_0$, and 4.9 for $A_2$. What a targeting program would act on is the share of workers above a threshold, and there the relocation is clearest: 0.5\% of bottom-quintile workers are in occupations above 0.5 under $A_2$, against 4.7\% under the benchmark, while in the top quintile the share rises from 34.2\% to 38.4\% (Table~\ref{tab:wages}).

\subsection{Sensitivity to the credit for integration-dependent exposure}
\label{sec:delta}

Almost the entire difference between the grounded labels and the benchmark lies in the E2/E3 class (Section~\ref{sec:label-distribution}), so the weight $\delta$ that Equation~\ref{eq:betadef} gives this class decides how large the difference looks. Table~\ref{tab:delta} recomputes the headline numbers for $\delta$ between 0, where only direct exposure counts, and 1, where integration-dependent exposure counts in full.

\begin{table}[t]
\centering
\small
\setlength{\tabcolsep}{4.5pt}
\begin{tabular}{lccccccc}
\toprule
& \multicolumn{3}{c}{Workers above 0.5 (millions)} & \multicolumn{3}{c}{Pearson $r$ with $R_k$} & Spearman $\rho$ \\
\cmidrule(lr){2-4}\cmidrule(lr){5-7}
$\delta$ & $B_0$ & $A_1$ & $A_2$ & $B_0$ & $A_1$ & $A_2$ & $A_2$ vs.\ $B_0$ \\
\midrule
0 (E1 only) & 5.5 & 13.9 & 8.6 & 0.511 & 0.675 & 0.689 & 0.54 \\
0.25 & 10.5 & 23.9 & 16.8 & 0.630 & 0.673 & 0.696 & 0.86 \\
0.5 (default) & 36.2 & 39.0 & 30.1 & 0.640 & 0.662 & 0.686 & 0.92 \\
0.75 & 61.5 & 46.0 & 38.0 & 0.613 & 0.646 & 0.670 & 0.92 \\
1 (E1 + E2/E3) & 74.1 & 48.7 & 45.0 & 0.582 & 0.630 & 0.651 & 0.92 \\
\bottomrule
\end{tabular}
\caption{Headline numbers under alternative weights $\delta$ for the E2/E3 class in Equation~\ref{eq:betadef}. Workers are U.S.\ employment in occupations with exposure above 0.5 \cite{bls_oews}. Pearson $r$ is against Anthropic's occupation-level exposure ($n=756$). The last column is the rank correlation between the calibrated index and the benchmark across occupations. Calibration weights are held at their fitted values. Benchmark values combine the released lower and upper bounds $\ell$ and $u$ as $\ell+\delta(u-\ell)$.}
\label{tab:delta}
\end{table}

Two conclusions survive every choice of $\delta$. Both context arms are more closely associated with observed occupational usage than the benchmark ($r$ of 0.630 to 0.696 against 0.511 to 0.640), and the gap is largest when only direct exposure counts. The level comparison does not survive. At $\delta\geq0.5$ the calibrated index counts fewer exposed workers than the benchmark, but at $\delta\leq0.25$ it counts more (8.6 against 5.5 million at $\delta=0$), because the grounded models credit direct exposure more often than GPT-4 did. The uncalibrated $A_1$ crosses later, counting more than the benchmark up to $\delta=0.5$ and fewer from $0.75$. Whether grounding \emph{lowers} measured exposure is therefore not a fact about AI capability but a consequence of how much unrealized integration counts today. Even the near-identical rank order of occupations depends on $\delta$ ($\rho=0.54$ at $\delta=0$). An index used to decide who receives reskilling support embeds this choice and should report it.

\section{Discussion}
\label{sec:discussion}

\paragraph{Measurement choices are normative choices.}
The largest single difference between the grounded index and the benchmark is not a disagreement about what AI can do. The grounded models credit direct exposure at least as often as GPT-4 did, but rarely credit a speed-up that would require software or integration that does not yet exist. How much such unrealized potential should count is a judgment about the purpose of the index, and Table~\ref{tab:delta} shows that it decides whether the index reports more or fewer exposed workers than the benchmark. Other choices have the same character: the rubric's counterfactual system, the rule that a task's time bottleneck sets its label, whether to anchor the index to one platform's usage at all, and a calibration objective that, on a target that is 92\% zero, rewards conservative models. None of these is wrong, but each moves exposure between occupations and across the wage distribution, and none can be settled by more evidence about AI. In line with the view that validity includes the consequences of using a score \cite{messick1995validity,jacobs2021measurement}, we suggest that an exposure index intended as a policy input should state these choices, report its headline numbers under the plausible alternatives, be inspectable at the level of the individual judgment, and be cheap enough to re-estimate that revision is routine.

\paragraph{What the results support.}
An ensemble of open-weight models separates tasks with observed usage from tasks without it better than the 2023 benchmark, with or without context, although the gain on all tasks is significant for only three of the four arms and partly reflects the ensemble's finer score resolution, although the gain within the tasks the benchmark rates feasible survives coarsening. Every arm separates physical from non-physical work better than the benchmark at full resolution, and both context arms are more closely associated with observed occupational usage than the benchmark for any credit $\delta$ given to integration-dependent exposure. Tested as a whole, the context mainly moves the level: it pulls down the exposure the models assign from their priors alone, and it does more than calibration to align the index with observed occupational usage, so an exposure measure can be improved without proprietary usage logs and $A_1$ is a complete index on its own. Calibration adds a modest but significant gain in occupation-level alignment, but by choosing a mixture of the models' existing levels rather than by matching the level of usage (Section~\ref{sec:calibration}).

\paragraph{Who is counted as exposed.}
At the default $\delta$, the calibrated index halves the benchmark's exposure for the lowest-paid fifth of workers and lowers it by 16\% for the highest-paid fifth, and most of this proportional steepening comes from calibration rather than from the context (Appendix~\ref{app:wages}). In absolute terms, the gap between the two fifths is nearly unchanged, because both ends are lowered. At the 0.5 threshold, the shift is clear: 0.5\% of the lowest-paid fifth are in occupations above 0.5 under $A_2$ against 4.7\% under the benchmark, while the share in the highest-paid fifth rises from 34.2\% to 38.4\%. This is where the anchor carries a risk of platform bias. Observed Claude usage is concentrated in computer, mathematical, and other knowledge work (Figure~\ref{fig:relocation-radar}, right), so a weight fit to its users' adoption favors models that score low wherever the platform is not used, whether or not AI could help there. Calibration thus imports the platform's user base into the level of the index, and a program that targets exposed occupations would direct more of its attention to higher-paid workers under $A_2$ than under $A_1$, and more under either than under the benchmark. The usage-free $A_1$ is the answer to this risk and should be the arm of choice where the aim is to describe capability rather than adoption. Even under $A_1$ the corpus, drawn from business and technology reporting, documents deployed systems in office work far better than in physical work, so a low score in physical and low-wage occupations should be read as an absence of documented present-day applicability, not as evidence that the work is safe from change.

\paragraph{Limitations.}
\label{sec:threats}
\emph{The behavioral anchor is one platform, and it is partly defined by the benchmark.} Penetration reflects adoption on Claude in one period, not what present systems can do, and the platform's users skew toward knowledge work, so the usage-anchored arm carries a platform bias that the uncalibrated arm avoids. \emph{The context is tested as a whole.} The no-context arms drop the passages, the descriptors, and the protocol together, so the design does not say which part of the context moves the labels. Arms that add one part at a time would settle this. \emph{The context shifts the level, not the ranking}, as the no-context arms rank tasks as well as or better than the context arms. \emph{The corpus is noisy}, Anglophone, and skewed toward business and technology outlets, so its coverage of physical and in-person work is thinner than its coverage of office work. \emph{The ensemble is small and sampled once.} The models have 14 to 32 billion parameters and judge the capability of larger frontier systems, which may make them conservative. We draw one sample per model and task, so within-model sampling variance is not measured. \emph{The weights are concentrated and the label space is coarse}, with half of the calibrated weight on the most conservative model and every judgment on a three-point grid. \emph{There is no external human validation of the labels.} The adjudication exercises run during development (Appendix~\ref{app:adjudication}) used curated item sets and occupation-level benchmark labels, so they cannot support a claim about which labels are correct. \emph{Exposure is not displacement.} What follows for workers depends on organizational choices, demand, regulation, and complementary skills.

\paragraph{Directions.}
A pre-registered comparison on randomly sampled contested tasks, labeled by people with experience in the occupations concerned, would replace the weakest validation, and anchoring to more than one platform would separate general adoption from what is specific to Claude.

\section{Conclusion}
\label{sec:conclusion}

This paper built an open, re-estimable task-level index of occupational AI exposure and used it to measure how much ``who is exposed'' depends on the choices inside the measure. The largest difference from the 2023 benchmark is not about what AI can do but about how much credit to give exposure that depends on integrations not yet built, a choice that no evidence settles and that decides whether the index counts more or fewer exposed workers than the benchmark. The open-weight ensemble ranks tasks better than the benchmark at full resolution and within the tasks the benchmark rates feasible, the context, tested as a whole, sets the level, and calibration to one platform's usage moves exposure, in proportional terms, further toward higher-wage occupations than the context does, which is why the index is released with and without it. An index used as a policy input should state such choices and report its numbers under the alternatives, so we release the framework as a scheduled pipeline, not a fixed dataset: the scores reported here describe one moment and are meant to be replaced.


\section*{Ethical Considerations Statement}

This work uses no individual-level data about workers. The occupational data are the public \onet database and public national employment statistics. The behavioral signal is the aggregated, task-level usage series released by Anthropic, which contains no conversation content and is gated for privacy at the source. The evidence corpus consists of news passages and research abstracts, and the released index carries passage identifiers and provenance rather than full articles. The human annotation summarized in Appendix~\ref{app:adjudication} involved three annotators not involved in the project and one author, who were informed of its purpose, worked only from task text and a rubric, and saw no personal or sensitive data. The measure describes where present AI systems could plausibly halve task time at equivalent quality. It is not a prediction of job loss. The main ethical risk of an exposure index is that it is read as a displacement forecast and used to justify decisions about individual workers. We therefore release the measure with its full decomposition and a dated vintage, so that an outdated score can be recognized as such.

\section*{Adverse Impact Statement}

An index that is more selective about which occupations count as exposed changes who is described as at risk. Under this measure, exposure moves away from clinical, physical, and interaction-heavy occupations toward document-intensive legal, financial, and analytical ones, and therefore up the wage distribution, so a program targeting exposed workers would direct more of its resources to higher-paid workers than one based on the benchmark. Under the calibrated arm 0.5\% of the lowest-paid fifth of workers are in occupations above 0.5, against 4.7\% under the benchmark, and in proportional terms most of that shift comes from calibration to observed Claude usage, whose users skew toward high-wage knowledge work. The uncalibrated arm $A_1$ avoids this platform bias, and we recommend it where the aim is to describe capability rather than adoption. If the measure is wrong, workers in the occupations it raises may face unwarranted concern, and those in the occupations it lowers may be overlooked by reskilling programs. The count of exposed workers also depends on the credit given to exposure that requires integrations not yet built (Section~\ref{sec:delta}). The corpus is Anglophone and drawn mostly from business and technology outlets, so the measure is least supported by documented evidence in the physical occupations it revises down most. The taxonomy is the U.S.\ \onet and the behavioral anchor is one platform, so transfer to other labor markets is not established. The index should be used as a task-level, versioned input to analysis rather than as an occupation-level verdict.

\section*{Generative AI Usage Statement}

Large language models are the measurement instrument in this work (Sections~\ref{sec:method} and~\ref{sec:setup}, prompts in Appendix~\ref{app:prompts}). Separately, Claude Code was used to write and run the stored-output scripts for the sensitivity and cross-fitting analyses (Section~\ref{sec:delta}, Appendix~\ref{app:sensitivity}). AI assistants were otherwise used for copy-editing and to accelerate analysis and figure code. The authors reviewed and verified all AI-assisted text, code, and numbers, and take full responsibility for the content of this paper. The research questions, design, data, methods, and conclusions are their own.

\bibliographystyle{ACM-Reference-Format}
\bibliography{references}

\appendix
\section{Prompt Construction}
\label{app:prompts}

This appendix gives the exact text sent to the inference models. Section~\ref{sec:method} describes what these components do, and this appendix shows what they contain, so that readers can reproduce them or criticize them on their merits.

\subsection{How a prompt is assembled}
\label{app:prompt-format}

Every occupation--task pair is sent as a two-message chat conversation. The system message is fixed across all pairs and all models and carries the exposure rubric. The user message is assembled per pair from the \onet fields, the retrieved evidence, and a fixed instruction block:

\begin{small}
\begin{verbatim}
[system]  <exposure rubric, Appendix A.2, identical for every pair>

[user]    Occupation:
          {occupation title}
          Occupation description:
          {occupation description, or "(No occupation description provided.)"}

          IWA (Intermediate Work Activity): {iwa title}

          DWA (Detailed Work Activity): {dwa title}

          Task:
          {task statement}

          NEWS:
          {retrieved passage block, or "(No NEWS context retrieved.)"}

          ONET SURVEY:
          {descriptor panel, or "(No O*NET survey context retrieved.)"}

          Instructions:
          <reasoning protocol and output contract, Appendix A.3>
\end{verbatim}
\end{small}

The IWA and DWA blocks are omitted when the field is empty. In the no-context arm, the \texttt{NEWS} and \texttt{ONET SURVEY} blocks are both dropped and a shorter instruction block is used instead. That arm therefore sees the occupation, description, work activities, and task, and nothing else. Passages are numbered in order of decreasing cross-encoder relevance and separated by a horizontal rule, each as \texttt{SOURCE n / Title: \{title\} / Content: \{text\}}. The source numbers are the only values a model may cite in \texttt{supporting\_sources}. The six \onet descriptors are shown as a fixed panel with the survey value and its response category side by side. A missing descriptor is shown as unavailable rather than dropped, so the panel always has six rows:

\begin{small}
\begin{verbatim}
O*NET Survey Results:
Workers in this occupation answered O*NET survey questions.
Important: these are occupation-level aggregates (job-level priors), not
task-level measurements for the specific task below.
Use this panel to identify potential substitution frictions, but do not let
it override the bottleneck described by the Task/IWA/DWA text.
Required descriptor panel:
- Survey question: "How frequently does your job require face-to-face
  discussions with individuals and within teams?" -> survey value: 87.5
  interpreted as "Every day".
- Survey question: "How much responsibility is there for the health and
  safety of other workers?" -> survey value: 61.2 interpreted as
  "High responsibility".
- ... four further descriptors ...
\end{verbatim}
\end{small}

The six survey questions and their response scales are:
\begin{itemize}
    \item face-to-face discussion (100 every day, 75 once a week or more, 50 once a month or more, 25 once a year or more, 0 never),
    \item responsibility for the health and safety of others (100 very high to 0 none),
    \item how automated the job is (100 completely to 0 not at all),
    \item importance of being exact or accurate (100 extremely important to 0 not important),
    \item importance of continuous, repetitive physical or mental activity (same scale as the previous item),
    \item frequency of written letters and memos (same scale as face-to-face discussion).
\end{itemize}

\subsection{The system prompt}
\label{app:system-prompt}

The system message is identical for every pair and every model. It states the counterfactual system, the counterfactual test, and the four exposure classes.

\begin{lstlisting}[basicstyle=\ttfamily\scriptsize, breaklines=true, breakindent=0pt, breakautoindent=false, columns=fullflexible, keepspaces=true, xleftmargin=0pt, frame=none, aboveskip=4pt, belowskip=4pt]
You are an expert AI labor economist evaluating occupational task exposure to advanced AI systems as of 2026. Your goal is to classify a task into one of four exposure categories (E0, E1, E2, E3) using the rubric below. You must reason carefully and follow the taxonomy strictly. Do not speculate beyond the rubric.

E Exposure Taxonomy
Consider the most capable LLM models available in 2026 in an agentic chatbot form. This system combines a frontier LLM with built in tool use, meaning it can make multi step plans, call tools, and take actions inside a web browser as well as inside common digital workspaces such as Microsoft 365 and Google Workspace when the worker has access and permissions set up. The system can navigate websites, fill forms, draft and edit documents, manipulate spreadsheets, create presentations, manage email and calendars, and execute structured workflows within these general purpose tools. Accessing internal systems through standard browser interaction using the worker's credentials counts as allowed capability, provided no specialized API level integration or custom enterprise tooling is required.
The system can retrieve up to date facts from the public internet via browsing, but it still depends on what it is allowed to access. It does not have access to private systems without credentials, proprietary databases unless provided, physical world sensors or actuators, or embedded enterprise back end integrations unless explicitly assumed.
Assume you are a worker with an average level of expertise in your role trying to complete the given task. You have access to this agentic chatbot and any other existing software or computer hardware tools mentioned in the task. You also have access to any commonly available technical tools accessible via a laptop, for example a microphone, speakers, a camera, standard office software, and common web services. You do not have access to any other physical tools or materials beyond what the task explicitly provides. Please label the given task according to the taxonomy below. When determining exposure, apply a counterfactual test: can access to the described agentic system reduce the time required to complete the task at equivalent quality by at least half? Focus on the primary bottleneck of the task rather than marginal improvements in small subcomponents.

E0: No exposure -> Label tasks E0 if direct access to the 2026 agentic chatbot interface cannot reduce the time it takes to complete this task with equivalent quality by half or more. Tasks should also be classified as E0 if more than half of the work is inherently physical, embodied, or depends on in person human interaction that cannot be substituted by text, workspace actions, or browser based actions. Examples include operating machinery, performing a physical inspection or repair that requires touch or specialized instruments not available through the laptop, delivering an in person demonstration, or physically presenting to stakeholders where real time interpersonal dynamics and embodiment are central. E0 also covers tasks where the binding constraint is access to physical environments, regulated sign off that must be done in person, or real time interpersonal dynamics such as speeches or instruction giving that cannot be meaningfully accelerated by drafting, planning, or digital coordination alone. Drafting materials for such activities may be E1, but performing the embodied activity itself remains E0 if the embodied component is the dominant share of time and effort. Although E0 is dominated by physical components, a task is still considered E0 it a chatbot is useful but cannot reduce completion time by at least 50% or more - i.e., a chatbot can be useful but will not be able to complete the task alone.

E1: Direct exposure -> Label tasks E1 if using a 2026 agentic chatbot, without any additional specialized enterprise integrations beyond standard browser and workspace access, can cut the time to complete the task at equivalent quality by at least half.
E1 assumes the agent operates through general purpose capabilities: text generation, reasoning, multi step planning, browsing public or credential accessible websites, and interacting with standard productivity software. It does not assume custom APIs, deep ERP or CRM embedding, structured database level querying, or organization specific tooling built specifically for the agent.
E1 typically covers text and code generation or editing, translation, summarization and question answering over provided or web accessible documents, structured web research with citations, and routine digital administrative work that is mostly browser or workspace based such as filling forms, scheduling, building a deck from notes, cleaning a spreadsheet, or drafting and sending messages with user approval.
E1 also includes tasks that benefit substantially from structured reasoning, cognitive scaffolding, and coordination support without requiring specialized enterprise integrations. This includes breaking down ambiguous problems into structured subcomponents and actionable plans, designing one time project roadmaps, timelines, milestone plans, execution checklists, drafting project briefs, outlining negotiation strategies, generating decision matrices, performing comparative analysis across documented options, synthesizing stakeholder inputs into coherent summaries, preparing structured reports, and orchestrating communication across teams using standard digital tools. If the worker could reasonably achieve a fifty percent or greater time reduction using only the agent's standalone capabilities and manual browser interaction, the task should be labeled E1.

E2: Exposure by LLM powered applications -> Label tasks E2 if having access to the 2026 agentic chatbot alone may not reduce the time it takes to complete the task by at least half in practice, but it is easy to imagine additional software, deeper integrations, or organization specific custom tooling built on top of the agent that would reduce the time it takes by half or more.
The key distinction from E1 is that in E2 the primary bottleneck is not general reasoning, drafting, or manual browser navigation. Instead, the bottleneck lies in reliable, structured, permission aware access to proprietary systems; automated interaction with internal databases; long horizon monitoring; compliance constrained workflows; high assurance environments; or tight coupling with internal business logic. Anything that can't be done with the tool out of the box.
Examples include searching over an organization's internal knowledge and systems such as data warehouses, ticketing tools, CRM, ERP, codebases, or logs and returning precise, permission aware answers; automating multi party workflows that require approvals, compliance checks, and policy enforcement; coordinating structured procurement or contracting systems that require formal submission pipelines; or providing real time assistance embedded directly in enterprise interfaces such as a support console pulling live customer records. In these cases, the standalone agent may assist partially through drafting or reasoning, but it cannot by itself remove the core system integration bottleneck. However, with deeper enterprise integration built on top of the agent, time savings of at least half are plausible.
Examples of software built on top of the agent that may help complete worker activities include tools for a home goods company that quickly process and summarize up to date internal sales and logistics data to inform product or marketing decisions; tools that suggest live responses for customer service agents while automatically pulling relevant account context; tools for legal work that aggregate and summarize prior internal case archives while retrieving the most recent controlling authority; or systems that automatically check prescriptions against contraindication databases and regulatory rules before submission. In each case, the decisive acceleration depends on structured integration beyond what a generic browser based agent can reliably accomplish alone.

E3: Exposure given image capabilities -> Suppose you had access to both the 2026 agentic chatbot and an integrated system capable of viewing, captioning, and generating images, including reading scanned PDFs, extracting text from images with high accuracy, interpreting diagrams, and analyzing video inputs. This system cannot reliably recover extremely fine physical measurements or tolerances unless those values are explicitly provided in text.
Label tasks as E3 only when access to these image capabilities, in addition to the agentic chatbot, enables a time reduction of at least fifty percent and visual understanding is a binding constraint for that reduction. E3 should be used only when visual interpretation, visual extraction, or image generation is central to the acceleration. If comparable time savings could be achieved using text alone, the task should not be labeled E3.
Examples include extracting text from non machine readable scanned documents, structuring data from screenshots, interpreting diagrams not available in textual form, reviewing visual layouts, annotating design mockups, or creating and editing digital images and figures according to specifications. If both enterprise integration and image capabilities are relevant, assign the label corresponding to the minimal additional capability required to achieve the time reduction. E3 applies only when image understanding is the critical enabler.
\end{lstlisting}

\subsection{The instruction block}
\label{app:instructions}

The instruction block closes the user message. It requires a subtask decomposition and an explicit bottleneck judgment rather than a bare label, and this is what the groundedness and faithfulness evaluator scores. It also makes an absent or irrelevant evidence bundle a legitimate outcome rather than an error.

\begin{lstlisting}[basicstyle=\ttfamily\footnotesize, breaklines=true, breakindent=1em, breakautoindent=true, columns=fullflexible, keepspaces=true, xleftmargin=0pt, frame=none, aboveskip=4pt, belowskip=4pt]
You are evaluating AI exposure for ONE specific job task.

Inputs you always receive:
- Occupation + occupation description
- Task, IWA (Intermediate Work Activity), DWA (Detailed Work Activity)
- O*NET SURVEY context: occupation-level metadata about typical work conditions (treat as background metadata, not task-level measurements)

Inputs you sometimes receive:
- NEWS context (SOURCE 1..N): retrieved news articles and research providing evidence of AI capabilities and technical feasibility. This section may be absent or marked "(No NEWS context retrieved.)" when no relevant documents were found for this task - this is expected and does not affect the validity of your assessment.

Reasoning procedure (follow in order):
1) Subtask decomposition:
   Break the task into 2-5 concrete subtasks that a worker would actually perform. Use the Task, IWA, and DWA text plus the O*NET metadata (work conditions, automation level, face-to-face requirements, etc.) to inform what the subtasks realistically involve. For each subtask, determine its exposure level (E0/E1/E2/E3). If NEWS sources are provided, use them as evidence for what AI can technically do and cite source numbers where applicable.

2) Overall task exposure: Synthesize the subtask assessments into ONE final label for the whole task. The overall label should reflect the task's bottleneck: the subtask(s) that dominate the worker's time determine the overall exposure.

Rules:
- O*NET survey data is job metadata - use it to understand the work environment and realistic constraints, not as a direct signal for exposure.
- If NEWS context is absent or not relevant to this specific task, set `relevant_context` to false and `supporting_sources` to [], and rely entirely on your expert reasoning.
- `supporting_sources` may only reference NEWS source numbers (SOURCE X). Do not cite sources that do not support the conclusion.
- Keep reasoning to one paragraph that covers the subtask breakdown, per-subtask assessment, and overall conclusion.

Return exactly one JSON object and nothing else:
{"reasoning (1 paragraph)": "<single paragraph covering subtask decomposition, per-subtask exposure, and overall assessment>", "relevant_context": true or false, "supporting_sources": [...], "exposure_label": "E0" or "E1" or "E2" or "E3"}
\end{lstlisting}

The no-context arm uses the much shorter block below instead. It asks for the label and a one-paragraph rationale, as the context arm does, without the evidence and descriptor rules. Together with the removal of the \texttt{NEWS} and \texttt{ONET SURVEY} blocks, this is the only difference between the two arms at the prompt level:

\begin{small}
\begin{verbatim}
You are given an occupation, occupation description, a task, and intermediate
and detailed work activity.
No NEWS context and no O*NET SURVEY context are provided in this run.

Task:
- Predict the extent to which this task is exposed to advanced AI systems
  (E0/E1/E2/E3).
- Keep reasoning to one paragraph that covers your assessment and overall
  conclusion.

Return exactly one JSON object and nothing else:
{"reasoning": "<single paragraph>", "exposure_label": "E0" or "E1" or "E2" or "E3"}
\end{verbatim}
\end{small}

\FloatBarrier
\section{Corpus Queries}
\label{app:corpus}

\subsection{The GDELT news query}
\label{app:gdelt-query}

News records are pulled month by month from the GDELT 2.0 Global Knowledge Graph on BigQuery \cite{leetaru2013gdelt}, against the partitioned table \texttt{gdelt-bq.gdeltv2.gkg\_partitioned}. The query returns metadata only, since GDELT does not store article text. The URLs it returns are passed to the \texttt{newspaper4k} scraper, which sometimes fails when a site is behind a paywall or firewall.
\begin{small}
\begin{verbatim}
SELECT
  DATE, SourceCommonName, SourceCollectionIdentifier, DocumentIdentifier,
  V2Themes, V2Persons, V2Organizations, V2Locations, V2Tone,
  Quotations, AllNames, Amounts
FROM `gdelt-bq.gdeltv2.gkg_partitioned`
WHERE DATE(_PARTITIONTIME) >= DATE('{start_date}') AND DATE(_PARTITIONTIME) <= DATE('{end_date}')
  -- English only, no translated records
  AND SourceCollectionIdentifier = 1 AND (TranslationInfo IS NULL OR TranslationInfo = '')
  AND GKGRECORDID NOT LIKE '%-T%'
  -- Source allowlist: one LIKE per domain, OR-ed together (Table B.1)
  AND ( LOWER(SourceCommonName) LIKE '%reuters.com%'
     OR LOWER(SourceCommonName) LIKE '%apnews.com%'
     OR ... )
  -- AI, automation, and emerging technology
  AND (  V2Themes LIKE '%ARTIFICIAL_INTELLIGENCE%' OR V2Themes LIKE '%AUTOMATION%'
      OR V2Themes LIKE '%TECH_AUTOMATION%'         OR V2Themes LIKE '%SOC_EMERGINGTECH%'
      OR AllNames LIKE '%ChatGPT%'   OR AllNames LIKE '%GPT-%'     OR AllNames LIKE '%Claude%'
      OR AllNames LIKE '%Gemini%'    OR AllNames LIKE '%Copilot%'  OR AllNames LIKE '%OpenAI%'
      OR AllNames LIKE '%Anthropic%' OR AllNames LIKE '%DeepSeek%'
      OR AllNames LIKE '%large language model%'  OR AllNames LIKE '%generative AI%'
      OR AllNames LIKE '%LLM%'       OR AllNames LIKE '%agentic AI%'  OR AllNames LIKE '%AI agent%'
      OR AllNames LIKE '%machine learning%'      OR AllNames LIKE '%natural language processing%'
      OR AllNames LIKE '%computer vision%'       OR AllNames LIKE '%predictive analytics%'
      OR AllNames LIKE '%robotic process automation%' OR AllNames LIKE '%autonomous vehicle%'
      OR AllNames LIKE '%neural network%'        OR AllNames LIKE '%deep learning%' )
\end{verbatim}
\end{small}

The query is deliberately broad at acquisition, because task-specific precision comes later from retrieval and reranking. The allowlist (Table~\ref{tab:allowlist}) is Anglophone and skewed toward business and technology reporting. This suits evidence about AI deployment, but it limits coverage of other labor markets.

\begin{table}[htbp]
\centering
\small
\begin{tabular}{p{0.24\linewidth} p{0.70\linewidth}}
\toprule
Tier & Domains \\
\midrule
Wire services and broadsheets &
reuters.com, apnews.com, bloomberg.com, wsj.com, ft.com, nytimes.com,
washingtonpost.com, theguardian.com, bbc.com, npr.org, economist.com, cnbc.com,
cnn.com, nbcnews.com, abcnews.go.com, cbsnews.com, usatoday.com, aljazeera.com,
scmp.com, theglobeandmail.com \\
\addlinespace
Technology and industry &
techcrunch.com, wired.com, arstechnica.com, technologyreview.com, theverge.com,
venturebeat.com, zdnet.com, techradar.com, computerweekly.com, siliconangle.com,
theregister.com \\
\addlinespace
Science and research &
nature.com, sciencedaily.com, newscientist.com, theconversation.com \\
\addlinespace
Business, general and government &
hbr.org, businessinsider.com, fortune.com, time.com, newsweek.com, thehill.com,
independent.co.uk, nypost.com, foxnews.com, yahoo.com, finance.yahoo.com,
govtech.com \\
\bottomrule
\end{tabular}
\caption{The 47 domains in the GDELT source allowlist. Matching is a case-insensitive substring test against \texttt{SourceCommonName}.}
\label{tab:allowlist}
\end{table}

\subsection{The Semantic Scholar query}
\label{app:s2-query}

Research abstracts are pulled from the Semantic Scholar bulk paper-search endpoint \cite{Kinney2023TheSS}. Four boolean queries are issued for every year from 2020 onward. A vertical bar means OR, and a leading plus sign makes a term required, so each query requires at least one technology term and at least one labor term.

\begin{small}
\begin{verbatim}
GET https://api.semanticscholar.org/graph/v1/paper/search/bulk

query 1: ("artificial intelligence" | AI | "machine learning")
           +("labor market" | employment | workforce | jobs)
query 2: ("generative ai" | LLM | ChatGPT | "agentic AI")
           +(displacement | automation | jobs | employment)
query 3: automation +("wage inequality" | unemployment
           | "occupational exposure" | "task displacement" | "white collar")
query 4: robots +("labor market" | "future of work" | employment | wages)

parameters:
  fields            = paperId, title, year, publicationDate, authors, abstract
  year              = {target year}
  sort              = publicationDate:desc
  publicationTypes  = JournalArticle, Review
  fieldsOfStudy     = Economics, Business, Sociology, Computer Science
\end{verbatim}
\end{small}

\FloatBarrier
\section{Inference and Retrieval Configuration}
\label{app:config}

Table~\ref{tab:model-pool} lists the ensemble, Table~\ref{tab:decoding} gives the decoding and serving settings, and Table~\ref{tab:retrieval-config} the retrieval hyperparameters. All were fixed before any evaluation against Anthropic penetration.

\begin{figure}[htbp]
    \centering
    \includegraphics[width=0.92\linewidth]{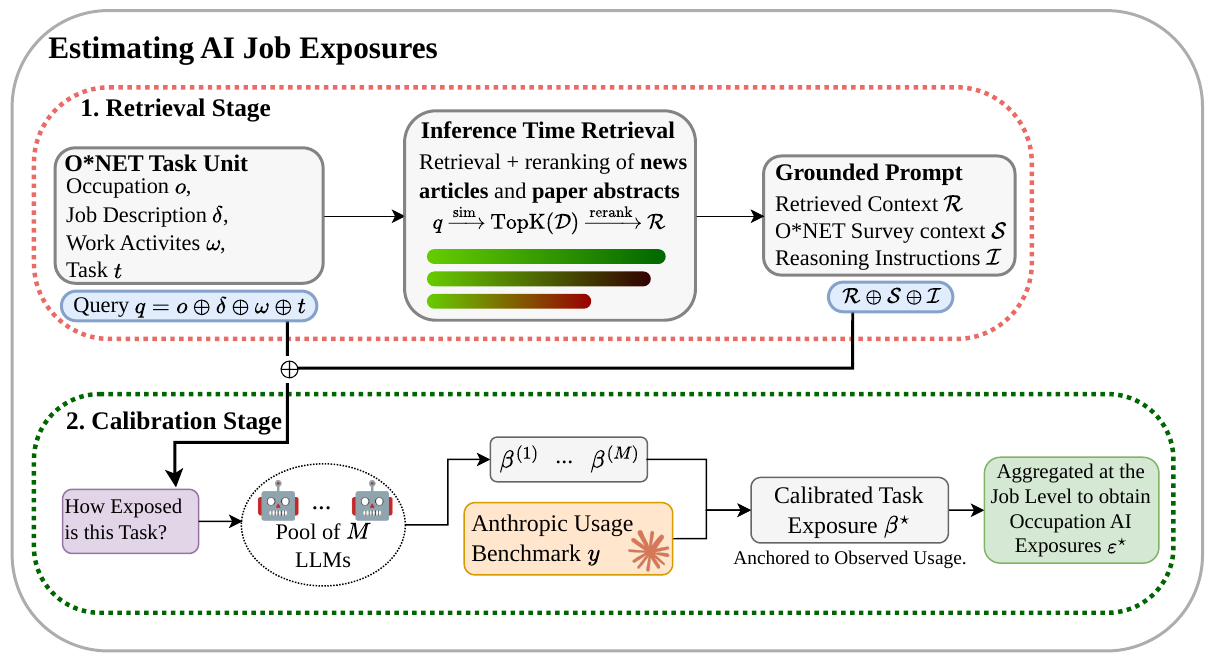}
    \caption{The evidence-grounded exposure pipeline, from an \onet occupation--task pair through corpus construction, hybrid retrieval and reranking, grounded multi-model inference, behavioral calibration, and task-share-weighted aggregation to an occupation-level score.}
    \Description{Flow diagram in two stages. In the retrieval stage, an O*NET task unit (occupation, job description, work activities, and task) forms a query that retrieves and reranks news articles and paper abstracts. The retrieved context, the O*NET survey context, and the reasoning instructions form a grounded prompt. In the calibration stage, a pool of M language models labels the task, their scores are combined into a calibrated task exposure anchored to the Anthropic usage benchmark, and task scores are aggregated to an occupation-level exposure.}
    \label{fig:pipeline}
\end{figure}

\paragraph{Hybrid retrieval.}
\label{app:retrieval}
The unit of retrieval is a passage. Abstracts and short articles are kept whole, and longer articles are split with a semantic sentence splitter \cite{kamradt2024semantic,chroma2024evaluating_chunking} into chunks of at least 800 characters (Table~\ref{tab:retrieval-config}). The corpus yields an index of over 176,000 passages. Every passage $c_j$ is encoded with BGE-M3 \cite{chen2024m3}, which produces a dense representation $\mathbf{d}_j\in\mathbb{R}^{1024}$ and a learned sparse lexical representation $\mathbf{s}_j$. Both are stored in an exact FAISS flat $L_2$ index \cite{douze2024faiss,johnson2017billion} together with the passage text and provenance. The query $q_{k,i}$ for a pair concatenates the occupation title, its description, the IWA and DWA, and the task statement, so that retrieval is conditioned on the occupation and not on the task string alone. We first retrieve the $P=40$ passages nearest to the dense query vector. We then give each candidate a hybrid score that combines dense proximity with sparse lexical overlap \cite{bruch2023analysis}:
\begin{equation}
r_{k,i,j}
=
\lambda
\max\!\left\{
0,
1-
\frac{
\left\lVert\mathbf{d}(q_{k,i})-\mathbf{d}_j\right\rVert_2^2
}{\sqrt{2}}
\right\}
+(1-\lambda)\,\mathbf{s}(q_{k,i})^{\top}\mathbf{s}_j,
\qquad
\lambda=0.75.
\label{eq:hybrid-relevance}
\end{equation}
A cross-encoder, the BGE reranker v2 M3 \cite{li2023making,nogueira2019passage}, then reranks the candidates. We keep the five with the highest scores and discard any whose hybrid score falls below a relevance threshold. Each model therefore receives between zero and five passages, and receiving none is a valid outcome rather than an error. The identifier of each passage is stored with the resulting assessment, so the evidence behind every judgment can be recovered.

\begin{table}[htbp]
\centering
\small
\setlength{\tabcolsep}{5pt}
\begin{tabular}{llll}
\toprule
Model & Developer & Parameters & Architecture \\
\midrule
DeepSeek-R1-Distill-Qwen-32B \cite{deepseek2025r1} & DeepSeek & 32B & Dense \\
Gemma~4 26B-A4B \cite{gemmateam2026gemma4} & Google & 26B total / 4B active & Mixture of experts \\
Ministral~3 14B \cite{liu2026ministral} & Mistral & 14B & Dense \\
Qwen3 30B-A3B \cite{yang2025qwen3} & Alibaba & 30B total / 3B active & Mixture of experts \\
GPT-OSS-20B \cite{openai2025gptoss120bgptoss20bmodel} & OpenAI & 20B & Mixture of experts \\
Nemotron-3-Nano-Omni 30B-A3B \cite{nvidia2026nemotron3nanoomni} & NVIDIA & 30B total / 3B active & Hybrid Mamba--Transformer mixture of experts \\
OpenReasoning-Nemotron 32B \cite{ahmad2025opencodereasoningiisimpletesttime,toshniwal2025genselect,moshkov2025aimo2winningsolutionbuilding} & NVIDIA & 32B & Dense \\
\bottomrule
\end{tabular}
\caption{The seven open-weight reasoning models in the ensemble. All run in their reasoning or thinking configuration, and within an arm every model sees identical prompts.}
\label{tab:model-pool}
\end{table}

\begin{table}[htbp]
\centering
\small
\begin{tabular}{ll}
\toprule
Setting & Value \\
\midrule
Temperature & 0.6 \\
Nucleus sampling ($p$) & 0.9 \\
Top-$k$ & 30 \\
Maximum generated tokens & 1{,}024 \\
Maximum context length, with-context arm & 8{,}900 tokens \\
Maximum context length, no-context arm & 2{,}800 tokens \\
Samples per model per task & 1 \\
Structured output & JSON schema \cite{dong2025xgrammar} \\
Serving framework & vLLM \cite{kwon2023efficient} \\
Hardware & 1 $\times$ NVIDIA H100 80GB per model \\
\bottomrule
\end{tabular}
\caption{Decoding and serving configuration, applied to every model in the ensemble and to both arms.}
\label{tab:decoding}
\end{table}

\begin{table}[htbp]
\centering
\small
\begin{tabular}{lll}
\toprule
Stage & Parameter & Value \\
\midrule
Chunking & Semantic breakpoint percentile & 80th \\
         & Minimum chunk size & 800 characters \\
         & Abstracts chunked & No, retained whole \\
         & News articles under 150 words & Not chunked \\
\midrule
Indexing & Embedding model & BGE-M3 \cite{chen2024m3} \\
         & Dense dimension & 1{,}024 \\
         & Index type & FAISS flat $L_2$, exact \cite{douze2024faiss} \\
         & Documents in corpus & 53{,}526 \\
         & Indexed passages & $\approx$176{,}000 \\
\midrule
Retrieval & Dense candidates $P$ & 40 \\
          & Hybrid dense weight $\lambda$ & 0.75 \\
          & Reranker & BGE reranker v2 M3 \cite{li2023making} \\
          & Passages after reranking & 5 \\
          & Passages shown to model & 0 to 5 \\
\bottomrule
\end{tabular}
\caption{Retrieval configuration. Queries are formed per occupation--task pair from the occupation title, the occupation description, the intermediate and detailed work activities, and the task statement.}
\label{tab:retrieval-config}
\end{table}

\paragraph{Compute and cost of a refresh.}
\label{app:compute}
Every model evaluates all 18,796 pairs, which gives $18{,}796 \times 7 = 131{,}572$ generations for the context arm and the same number for the no-context arm. Index construction (embedding roughly 176,000 passages and building the flat index) and cross-encoder reranking of 40 candidates per task ran on the same class of hardware. Retrieval is cached per pair and shared across all seven models, so the retrieval stage is paid once per update rather than once per model. Because the cache is compared with the previous run, a refresh re-runs inference only for the pairs whose evidence changed.

\paragraph{Response schema.}
\label{app:schema}
Generation is constrained to the following JSON schema, whose \texttt{reasoning} key overrides the longer key name shown in the prompt's example object. The upper bound on \texttt{supporting\_sources} is set per task to the number of passages actually shown, so a model cannot cite a source that does not exist. The no-context arm uses a reduced schema with only \texttt{reasoning} and \texttt{exposure\_label}.

\begin{small}
\begin{verbatim}
{
  "type": "object",
  "properties": {
    "reasoning":          {"type": "string", "maxLength": 2500},
    "relevant_context":   {"type": "boolean"},
    "supporting_sources": {"type": "array",
                           "items": {"type": "integer", "minimum": 1, "maximum": 5}},
    "exposure_label":     {"type": "string", "enum": ["E0", "E1", "E2", "E3"]}
  },
  "required": ["reasoning", "relevant_context", "supporting_sources", "exposure_label"],
  "additionalProperties": false
}
\end{verbatim}
\end{small}

\paragraph{Calibration splits.}
\label{app:splits}
Table~\ref{tab:splits} records the fold sizes. The two samples match slightly different task sets and were split independently, and 78.3\% of the with-context held-out tasks fall in the no-context training fold. To compare all arms on the same tasks without leakage, every task-level result uses the with-context held-out fold, restricted to the 3,391 tasks that also have a no-context and a benchmark score (257 positives). For these comparisons, the no-context weights are refitted on the matching training fold. Occupation-level scores for the no-context arm use the weights fitted on its own sample.

\begin{table}[htbp]
\centering
\small
\begin{tabular}{lrrrr}
\toprule
Arm & Matched tasks & Training & Held out & Held-out positives \\
\midrule
With context ($A_1$, $A_2$) & 16{,}956 & 13{,}564 & 3{,}392 & 258 \\
No context ($A_0$)          & 17{,}428 & 13{,}942 & 3{,}486 & 254 \\
\bottomrule
\end{tabular}
\caption{Fold sizes for the two calibration samples. Both are 80/20 splits stratified on whether penetration is positive, which holds the positive rate at 7.6\% in each fold. The held-out positive count for $A_0$ is reported over the tasks also present in the with-context sample.}
\label{tab:splits}
\end{table}

\paragraph{Samples, counts, and means that appear in the paper.}
\label{app:counts}
Several numbers that look alike refer to different samples. \onet 30.2 has 923 detailed occupations and 18,796 occupation--task pairs. A retrieval bundle was built for 18,764 of them, and every model scored all 18,796. Of the pairs, 18,205 have both a grounded score and a benchmark task label (the label-flow table in Appendix~\ref{app:flows} was built with a slightly different task match and covers 18,203), 18,200 also have a no-context score and form the sample for the mean task score in Table~\ref{tab:ablation-summary}, and 18,183 match the physical-task classification. At the task level, the pairs contain 16,956 distinct normalized task strings that match the Anthropic task file and for which all seven models returned a valid label. This is the calibration sample, and the no-context run, with its own set of valid responses, matches 17,428. The three means reported for the calibrated index are taken over three of these samples: 0.250 over the 16,956 calibration tasks, 0.257 over the 18,200 pairs, where a task recurs once for every occupation that contains it, and 0.245 over the 923 detailed occupations after weighting tasks by time shares. At the occupation level, the 923 detailed occupations collapse to 798 six-digit SOC groups, the granularity of Anthropic's occupation file. Of these, 756 match an Anthropic occupation score, and 750 have both a median wage and an employment count in the OEWS release used for the wage and worker counts.

\paragraph{Stability of the fitted calibration weights.}
\label{app:weight-stability}
Two checks test whether the single large weight reported in Section~\ref{sec:weights} is an artifact of one split. Across 300 bootstrap resamples of the training fold, the weights are stable. Their standard deviations are between $0.011$ and $0.014$, the largest weight has a 95\% interval of $[0.485, 0.540]$, and the two zeroed models receive zero weight in 81\% and 100\% of resamples. Re-fitting under alternative splits is more informative (Table~\ref{tab:split-sweep}). The $B_0$ column there is the occupation-averaged benchmark score rather than the task label used in Table~\ref{tab:ablation-summary}, which is why its AUC differs from 0.766. We hold the negatives at a 20\% test share and vary how many of the 1,288 positive tasks remain available for training. The largest weight then rises monotonically from $0.497$ with 1,159 training positives to $0.633$ with 388, a drift far larger than the bootstrap standard deviation. Balancing the training classes by downsampling zeros drives it to $0.03$, with the ensemble close to uniform. Held-out ranking is unaffected throughout, with AUC between $0.789$ and $0.794$, against $0.779$ to $0.785$ for the occupation-averaged benchmark score. The concentration is therefore a property of how many positive tasks the fit sees, not of one random draw.

\begin{table}[htbp]
\centering
\small
\setlength{\tabcolsep}{5pt}
\begin{tabular}{rrccc}
\toprule
Train & Test & \multicolumn{2}{c}{Held-out AUC} & Largest \\
\cmidrule(lr){3-4}
pos. & pos. & $A_2$ & $B_0$ & weight \\
\midrule
1{,}159 & 129 & 0.789 & 0.779 & 0.497 \\
1{,}030 & 258 & 0.791 & 0.781 & 0.516 \\
888 & 400 & 0.792 & 0.784 & 0.539 \\
688 & 600 & 0.794 & 0.784 & 0.573 \\
644 & 644 & 0.793 & 0.784 & 0.581 \\
388 & 900 & 0.793 & 0.785 & 0.633 \\
\bottomrule
\end{tabular}
\caption{Calibration re-fitted under alternative allocations of the 1,288 positive tasks, with negatives held at a 20\% test share. Each row averages 25 random draws, and the second row has the same allocation as the split used throughout. The $B_0$ column is the occupation-averaged \bench $\beta$, evaluated on the same fold as $A_2$ in every row.}
\label{tab:split-sweep}
\end{table}

\paragraph{Anthropic's automation factor.}
\label{app:anthropic}
The factor $\alpha_i$ in Equation~\ref{eq:anthropic-task-penetration} gives greater weight to uses Anthropic classifies as automative rather than augmentative:
\begin{equation}
\alpha_i
=
\frac{1}{2}
+
\frac{1}{2}
\left(
\frac{
\mathrm{ClaudeWorkUsage}_{i}\,\mathrm{AutoShare}_{i}
+\mathrm{APIUsage}_{i}
}{
\mathrm{ClaudeWorkUsage}_{i}+\mathrm{APIUsage}_{i}
}
\right),
\label{eq:anthropic-automation-factor}
\end{equation}
where $\mathrm{AutoShare}_{i}$ is the share of Claude.ai usage for task $i$ that is classified as automative. API traffic is treated as automative because it is associated with deeper workflow integration. A task observed only in augmentative interactions receives $\alpha_i=0.5$, and one observed only in automative interactions or API traffic receives a value approaching 1. The released task file contains 17,998 rows representing 17,992 unique task strings, of which 1,354 have positive penetration. After matching to the seven-model task sample, the calibration dataset contains 16,956 tasks, of which 1,288 (7.6\%) have positive penetration.

\FloatBarrier
\section{Additional Results}
\label{app:detailed-tables}

\subsection{Task-level discrimination}

\begin{figure}[htbp]
    \centering
    \includegraphics[width=0.7\linewidth]{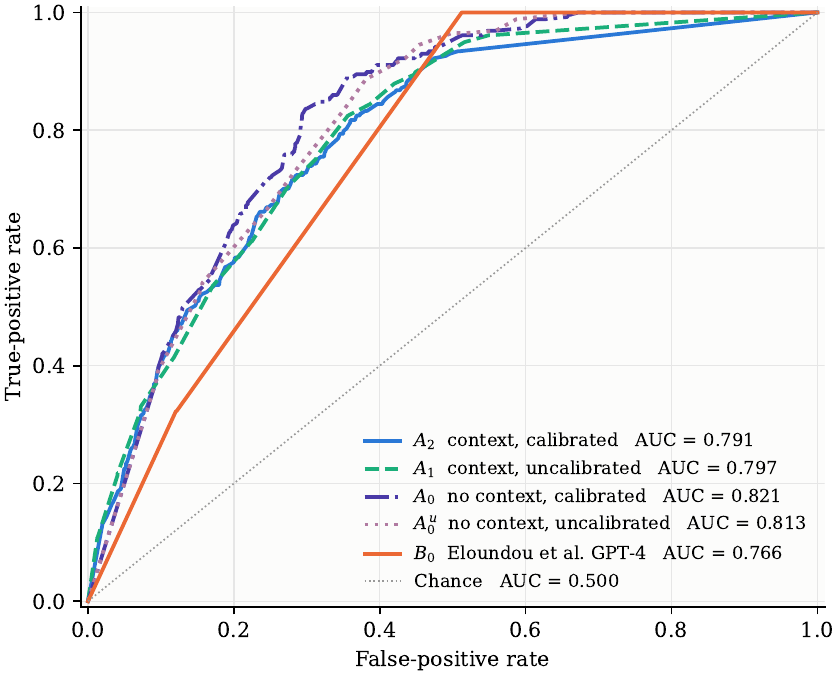}
    \caption{Held-out task-level ROC curves for detecting positive Anthropic task penetration, for all five arms on the common fold. The benchmark curve reaches a true-positive rate of 1.0 at a false-positive rate near 0.5. This is because the feasibility gate in Equation~\ref{eq:anthropic-task-penetration} means that no positive-penetration task can have $\beta^{\mathrm{E}}=0$, so the benchmark cannot produce a false negative on this target.}
    \Description{ROC curves for five measures. The four ensemble arms rise steeply and stay close to one another, with areas under the curve between 0.79 and 0.82. The benchmark curve is a piecewise straight line that reaches a true-positive rate of 1.0 at a false-positive rate near 0.5, with an area under the curve of 0.766.}
    \label{fig:task_level_roc}
\end{figure}

Table~\ref{tab:task_auc} gives, for each arm, the held-out AUC and average precision for detecting tasks with positive Anthropic penetration, with paired bootstrap differences against the benchmark, and the AUC within the tasks that the benchmark rates feasible. Figure~\ref{fig:task_level_roc} shows the corresponding ROC curves. These results are summarized in Table~\ref{tab:ablation-summary} and discussed in Section~\ref{sec:task_auc}. Because $B_0$ takes three values and the ensemble scores many, part of the all-task difference is resolution. Rounded to the benchmark's grid (E0 up to 0.25, E1 from 0.75), the all-task AUCs of $A_0^{u}$, $A_0$, $A_1$, and $A_2$ are 0.771, 0.788, 0.765, and 0.748, none significantly different from the benchmark's 0.766 (paired bootstrap $p$ between 0.09 and 0.92), while their feasible-set AUCs are 0.598, 0.639, 0.618, and 0.626, all above the benchmark's 0.543 at $p<0.01$.

\begin{table}[htbp]
\centering
\small
\setlength{\tabcolsep}{4pt}
\begin{tabular}{lccccc}
\toprule
Measure & AUC & $\Delta$AUC vs.\ $B_0$ [95\% CI] & AP & $\Delta$AP vs.\ $B_0$ [95\% CI] & AUC, feasible \\
\midrule
$B_0$: \bench benchmark & 0.766 & -- & 0.152 & -- & 0.543 \\
$A_0^{u}$: No context, uncalibrated & 0.813 & $+0.047$ $[+0.024,+0.069]$ & 0.204 & $+0.053$ $[+0.029,+0.078]$ & 0.653 \\
$A_0$: No context, calibrated & 0.821 & $+0.055$ $[+0.032,+0.077]$ & 0.215 & $+0.063$ $[+0.039,+0.089]$ & 0.670 \\
$A_1$: Context, uncalibrated & 0.797 & $+0.031$ $[+0.007,+0.055]$ & 0.223 & $+0.071$ $[+0.041,+0.107]$ & 0.642 \\
$A_2$: Context, calibrated & 0.791 & $+0.025$ $[-0.001,+0.051]$ & 0.221 & $+0.069$ $[+0.039,+0.104]$ & 0.642 \\
\bottomrule
\end{tabular}
\caption{Held-out task-level discrimination of positive Anthropic penetration on the common fold ($n=3{,}391$ with 257 positives). AP is average precision, against a base rate of 0.076. The last column restricts the fold to the 1,863 tasks with $\beta^{\mathrm{E}}\geq0.5$, which contain every positive (base rate 0.138). Differences are paired bootstrap estimates over 5,000 resamples. All differences against $B_0$ are significant at $p<0.01$ except the AUC of $A_2$ ($p=0.055$).}
\label{tab:task_auc}
\end{table}

\subsection{Contrasts between arms}
\label{app:contrasts}

Table~\ref{tab:contrasts} isolates the effect of the context and of calibration by comparing arms that differ in only one of the two. It shows the pattern described in Section~\ref{sec:ablation-synthesis}: the context leaves the task-level ranking of usage almost unchanged and slightly lowers the ranking of physical tasks, but it improves occupation-level alignment with observed usage more than calibration does.

\begin{table}[htbp]
\centering
\small
\setlength{\tabcolsep}{3.5pt}
\begin{tabular}{lcccc}
\toprule
Contrast & $\Delta$AUC & $\Delta$AP & $\Delta r$ with $R_k$ & $\Delta$AUC \\
 & (usage) & (usage) & (occupations) & (physical) \\
\midrule
Context, uncalibrated ($A_1-A_0^{u}$) & $-0.015$ & $+0.018$ & $+0.059$ & $-0.031$ \\
 & {\scriptsize $[-0.034,+0.003]$} & {\scriptsize $[-0.006,+0.049]$} & {\scriptsize $[+0.036,+0.081]$} & {\scriptsize $[-0.035,-0.027]$} \\
\addlinespace
Context, calibrated ($A_2-A_0$) & $-0.030$ & $+0.006$ & $+0.051$ & $-0.048$ \\
 & {\scriptsize $[-0.051,-0.010]$} & {\scriptsize $[-0.022,+0.037]$} & {\scriptsize $[+0.027,+0.075]$} & {\scriptsize $[-0.052,-0.044]$} \\
\addlinespace
Calibration, no context ($A_0-A_0^{u}$) & $+0.008$ & $+0.010$ & $+0.032$ & $-0.001$ \\
 & {\scriptsize $[+0.002,+0.015]$} & {\scriptsize $[+0.006,+0.016]$} & {\scriptsize $[+0.017,+0.047]$} & {\scriptsize $[-0.002,-0.000]$} \\
\addlinespace
Calibration, context ($A_2-A_1$) & $-0.006$ & $-0.002$ & $+0.024$ & $-0.018$ \\
 & {\scriptsize $[-0.017,+0.005]$} & {\scriptsize $[-0.017,+0.012]$} & {\scriptsize $[+0.014,+0.035]$} & {\scriptsize $[-0.020,-0.016]$} \\
\bottomrule
\end{tabular}
\caption{Paired differences between arms that differ in context only or in calibration only, with bootstrap 95\% confidence intervals. Usage outcomes are on the common held-out fold, $r$ is the occupation-level Pearson correlation with $R_k$ ($n=756$), and the physical outcome is on all matched occupation--task pairs.}
\label{tab:contrasts}
\end{table}

\subsection{Occupation and industry level}

\begin{figure}[htbp]
    \centering
    \includegraphics[width=0.8\linewidth]{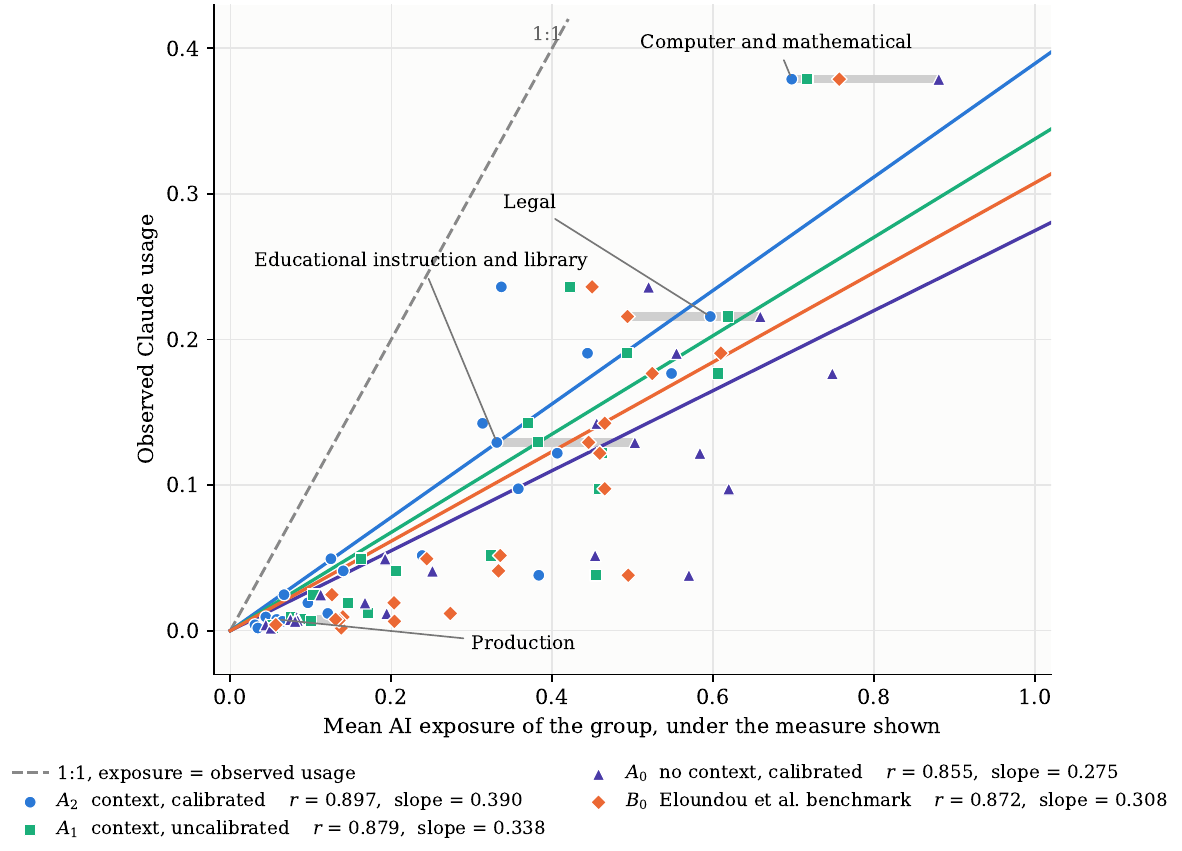}
    \caption{Observed Claude usage against mean group exposure across the 22 \onet major occupational groups, for four of the five arms. $A_0^{u}$ is omitted for readability; its group-level Pearson correlation is $0.802$, its Spearman correlation $0.898$, and its slope $0.223$. The solid lines are least-squares regressions through the origin, one per measure, and the dashed line marks equality between exposure and usage. Four groups are labeled, and the grey bar behind each joins that group's four exposure estimates at its single observed usage value. Slopes are $0.308$ for $B_0$, $0.275$ for $A_0$, $0.338$ for $A_1$, and $0.390$ for $A_2$. Group-level Pearson correlations are $0.872$, $0.855$, $0.879$, and $0.897$, and Spearman correlations are $0.848$, $0.892$, $0.910$, and $0.911$. With $n=22$, the intervals around these values are wide.}
    \Description{Scatter plot of observed Claude usage against mean exposure for 22 occupational groups, with one marker type and one regression line through the origin per arm. All four lines rise from the origin. The no-context line is the flattest and the calibrated grounded line is the steepest. Computer and mathematical occupations have the highest usage, near 0.38, and production occupations have usage near zero.}
    \label{fig:industry_alignment}
\end{figure}

Table~\ref{tab:correlation} reports the occupation-level correlations with Anthropic's observed exposure $R_k$ and the bootstrap differences against the benchmark discussed in Section~\ref{sec:occupation-alignment}. Figure~\ref{fig:industry_alignment} shows the same comparison for the 22 major occupational groups.

\begin{table}[htbp]
\centering
\small
\begin{tabular}{lccccc}
\toprule
Measure & Pearson $r$ & $\Delta r$ & 95\% CI & $p$ & Spearman $\rho$ \\
\midrule
$B_0$: \bench benchmark & 0.640 & -- & -- & -- & 0.701 \\
$A_0^{u}$: No context, uncalibrated & 0.602 & $-0.038$ & [-0.073, -0.001] & 0.040 & 0.690 \\
$A_0$: No context, calibrated & 0.635 & $-0.006$ & [-0.041, 0.031] & 0.776 & 0.688 \\
$A_1$: Context, uncalibrated & 0.662 & $+0.021$ & [-0.012, 0.054] & 0.214 & 0.700 \\
$A_2$: Context, calibrated & \textbf{0.686}$^{*}$ & \textbf{+0.046} & \textbf{[0.009, 0.081]} & \textbf{0.013} & \textbf{0.702} \\
\bottomrule
\end{tabular}
\caption{Occupation-level alignment with observed Anthropic exposure $R_k$ over the $n=756$ six-digit SOC groups matched to Anthropic occupation data, of the 923 scored. Pearson differences are computed relative to $B_0$ using 5,000 paired bootstrap resamples.}
\label{tab:correlation}
\end{table}

\subsection{Independent target}
\label{app:independent}

Table~\ref{tab:independent} gives the full results for the independent target of Section~\ref{sec:triangulation}, the physical-task classification of \citet{tomlinson2025workingwithai}. Rounded to the benchmark's three-point grid, the AUCs of $A_0^{u}$, $A_0$, $A_1$, and $A_2$ are 0.890, 0.842, 0.827, and 0.767, against 0.815 for the benchmark, with bootstrap intervals that exclude zero for every difference, and the calibrated context arm is the only one that falls below the benchmark once its resolution is removed.

\begin{table}[htbp]
\centering
\small
\setlength{\tabcolsep}{4pt}
\begin{tabular}{lccc}
\toprule
Measure & AUC & Scored $\leq 0.10$ & Mean score, physical tasks \\
\midrule
$B_0$: \bench benchmark & 0.815 & 79.3\% & 0.120 \\
$A_0^{u}$: No context, uncalibrated & 0.915 & 68.5\% & 0.134 \\
$A_0$: No context, calibrated & 0.914 & 87.4\% & 0.073 \\
$A_1$: Context, uncalibrated & 0.885 & 83.5\% & 0.061 \\
$A_2$: Context, calibrated & 0.866 & 89.8\% & 0.046 \\
\bottomrule
\end{tabular}
\caption{Independent target. The AUC measures how well each score separates non-physical from physical tasks (18,183 pairs, 42.4\% physical). The next two columns give the share of physical tasks scored at 0.10 or below and the mean score over the physical tasks.}
\label{tab:independent}
\end{table}

\subsection{Exposure by wage}
\label{app:wages}

Table~\ref{tab:wages} groups workers into employment-weighted quintiles of occupational median wage, from the May 2021 OEWS release \cite{bls_oews}, held fixed across all arms so that the counts differ only through the exposure scores, and reports the mean exposure of each arm in each quintile. All measures rise with wage, but the grounded and calibrated index rises most steeply in proportional terms. In absolute terms, the top-to-bottom gap is 0.32 under the benchmark, 0.37 under $A_1$, and 0.33 under $A_2$, because the index lowers both ends. Calibration steepens the proportional gradient more than the context does: the ratio of the top to the bottom quintile rises from 2.9 to 4.7 when the no-context arm is calibrated and from 3.9 to 4.9 when the context arm is, against 2.9 to 3.9 and 4.7 to 4.9 when the context is added to the uncalibrated and to the calibrated arm.

\begin{table}[htbp]
\centering
\small
\setlength{\tabcolsep}{4.5pt}
\begin{tabular}{lccccccc}
\toprule
Wage quintile & Median wage & Workers & $B_0$ & $A_0^{u}$ & $A_0$ & $A_1$ & $A_2$ \\
\midrule
1 (lowest) & \$24k--30k & 25.4M & 0.177 & 0.257 & 0.137 & 0.127 & 0.086 \\
2 & \$30k--38k & 26.0M & 0.309 & 0.398 & 0.303 & 0.251 & 0.205 \\
3 & \$38k--49k & 26.5M & 0.332 & 0.361 & 0.284 & 0.260 & 0.222 \\
4 & \$49k--77k & 26.0M & 0.437 & 0.619 & 0.511 & 0.413 & 0.349 \\
5 (highest) & \$77k--202k & 26.0M & 0.498 & 0.734 & 0.640 & 0.493 & 0.418 \\
\midrule
\multicolumn{3}{l}{Ratio of quintile 5 to quintile 1} & 2.8 & 2.9 & 4.7 & 3.9 & 4.9 \\
\bottomrule
\end{tabular}
\caption{Employment-weighted mean exposure by wage quintile, over the 750 six-digit occupations with wage and employment data (99.5\% of covered employment). Quintiles hold roughly equal numbers of workers. Under $A_2$, 0.5\%, 23.9\%, 21.9\%, 30.6\%, and 38.4\% of the workers in each quintile are in occupations above 0.5, against 4.7\%, 33.7\%, 25.2\%, 40.8\%, and 34.2\% under $B_0$.}
\label{tab:wages}
\end{table}

\subsection{Sensitivity of the headline numbers}
\label{app:sensitivity}

All analyses in this subsection are computed from the stored outputs of the main run, without new model inference. They reproduce the published occupation scores exactly at the default settings.

\paragraph{Weight on integration-dependent exposure.}
Table~\ref{tab:delta-full} extends Table~\ref{tab:delta} with employment-weighted mean exposure. For the ensemble, labels on the grid are remapped exactly. For the 2.26\% of tasks whose score is averaged over several occupations, the score is interpolated linearly between grid points. For the benchmark, the released lower and upper bounds $\ell$ and $u$ are combined as $\ell+\delta(u-\ell)$, which reproduces the published $\beta$ at $\delta=\tfrac{1}{2}$. The employment-weighted mean of the calibrated index lies above that of the benchmark when only direct exposure counts and below it for $\delta\geq0.25$, while the uncalibrated $A_1$ lies above it for $\delta\leq0.25$ and below it from $\delta=0.5$.

\begin{table}[htbp]
\centering
\small
\begin{tabular}{lccc}
\toprule
& \multicolumn{3}{c}{Employment-weighted mean exposure} \\
\cmidrule(lr){2-4}
$\delta$ & $B_0$ & $A_1$ & $A_2$ \\
\midrule
0 & 0.146 & 0.226 & 0.172 \\
0.25 & 0.249 & 0.268 & 0.214 \\
0.5 & 0.352 & 0.309 & 0.256 \\
0.75 & 0.454 & 0.350 & 0.298 \\
1 & 0.557 & 0.391 & 0.341 \\
\bottomrule
\end{tabular}
\caption{Employment-weighted mean occupational exposure under alternative weights $\delta$ for the E2/E3 class.}
\label{tab:delta-full}
\end{table}

\paragraph{Weighting of zero-penetration tasks.}
Equation~\ref{eq:linear-zero-weight} lowers the weight of zero-penetration tasks that the ensemble rates as exposed. Refitting with every task at full weight moves weight further toward Qwen3 (from 0.511 to 0.614) and lowers the mean task score from 0.250 to 0.239 and the number of workers in occupations above 0.5 from 30.1 to 26.2 million. Held-out AUC (0.797 against 0.795) and the occupation-level correlation with $R_k$ (0.686 against 0.687) are unchanged. Both objectives favor the most conservative model, as expected for a target that is 92\% zero.

\paragraph{Cross-fitting by occupation.}
We split the 798 six-digit occupations into five random folds. For each fold, the weights are fitted on all calibration tasks that do not occur in any occupation of that fold and are then applied to that fold's occupations. Across folds, the Qwen3 weight ranges from 0.508 to 0.525 and every other weight moves by less than 0.025. The cross-fitted index differs from the published one by at most 0.011 for any occupation, and its correlation with $R_k$ is 0.686, against 0.640 for the benchmark ($\Delta r=+0.045$, 95\% CI $[0.010, 0.080]$, $p=0.013$, 5,000 paired bootstrap resamples).

\FloatBarrier
\section{Label Flows Between the Benchmark and the Grounded Measure}
\label{app:flows}

\begin{figure}[htbp]
    \centering
    \includegraphics[width=0.85\linewidth]{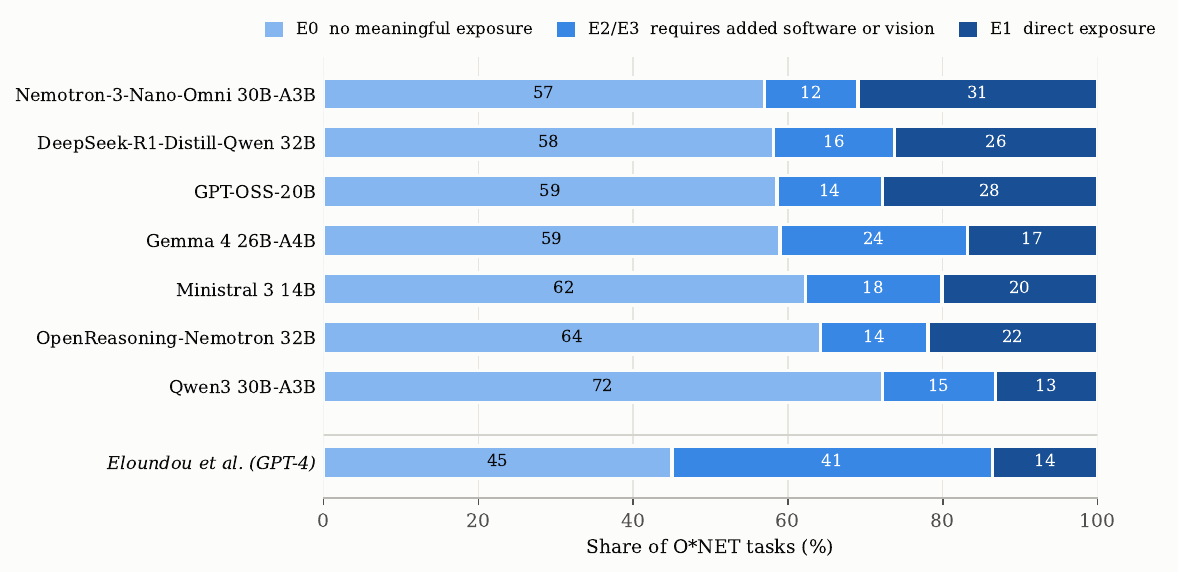}
    \caption{Distribution of exposure labels across the 16,956 normalized \onet tasks in the calibration sample, for all seven grounded models and the \bench GPT-4 labels, ordered by E0 share. The grounded models place more tasks in E0 and far fewer in E2/E3 (12\% to 24\%, against 41\% for the benchmark), while assigning E1 at comparable or higher rates.}
    \Description{Horizontal stacked bar chart with one bar per model and one for the benchmark. Each bar splits 100 percent of tasks into E0, E2/E3, and E1. The seven grounded models assign 57 to 72 percent of tasks to E0, 12 to 24 percent to E2/E3, and 13 to 31 percent to E1. The benchmark assigns 45 percent to E0, 41 percent to E2/E3, and 14 percent to E1.}
    \label{fig:exposure_dist}
\end{figure}

This appendix expands Section~\ref{sec:label-distribution}. We compare the two label sets over the 18,203 occupation--task pairs for which both a benchmark label and a grounded label exist. The calibrated score is mapped back onto the E-label grid, with E0 up to 0.25, E1 from 0.75, and E2/E3 in between.

\begin{table}[htbp]
\centering
\small
\setlength{\tabcolsep}{5pt}
\begin{tabular}{lrrrr rrr}
\toprule
& \multicolumn{4}{c}{Grounded label, $n$ (row \%)} & \multicolumn{3}{c}{Nonzero Claude penetration (\%)} \\
\cmidrule(lr){2-5}\cmidrule(lr){6-8}
Benchmark label & E0 & E2/E3 & E1 & Total & E0 & E2/E3 & E1 \\
\midrule
E0 (no exposure)      & 7{,}866 (97.9) & 157 (2.0) & 13 (0.2) & 8{,}036 & 0.1 & 0.0 & 0.0 \\
E1 (direct)           & 535 (21.2) & 1{,}099 (43.5) & 892 (35.3) & 2{,}526 & 8.6 & 16.2 & 29.6 \\
E2 (tool-dependent)   & 3{,}149 (41.2) & 3{,}117 (40.8) & 1{,}375 (18.0) & 7{,}641 & 12.5 & 13.6 & 26.2 \\
\midrule
Total                 & 11{,}550 & 4{,}373 & 2{,}280 & 18{,}203 & 3.9 & 13.8 & 27.4 \\
\bottomrule
\end{tabular}
\caption{Contingency between the benchmark task label and the calibrated grounded label, with the share of pairs in each cell that have nonzero Anthropic task penetration. The E0 row of the right block is mechanically near zero, because the feasibility gate forces penetration to zero for any task the benchmark rated E0.}
\label{tab:label_flows_full}
\end{table}

Two things are visible in Table~\ref{tab:label_flows_full}. First, the grounded measure is far more likely to move a task down than up. The benchmark's E2 category, which covers 42.0\% of matched pairs, is split almost evenly between E0 and E2/E3, with 18.0\% raised to E1. In contrast, only 2.1\% of benchmark E0 tasks move up at all. Overall, 4,524 of the 6,328 label changes (71.5\%) start in the benchmark's E2 row. Second, the direction of a move predicts observed usage within the two benchmark classes where this can be checked. Reading across the E1 row, the share of tasks with nonzero penetration falls from 29.6\% for the tasks we keep at E1, to 16.2\% for those we lower to E2/E3, and to 8.6\% for those we lower to E0. The E2 row is monotone in the same direction. The benchmark label alone cannot make these distinctions, since every task in a row has the same benchmark value. The E0 row cannot be validated in this way, because penetration is gated on $\beta^{\mathrm{E}}\geq0.5$. The 170 pairs that the grounded measure raises out of E0 therefore cannot be verified against this target, which is one reason why Section~\ref{sec:physical} uses an external classification instead.

The same table shows the measure's false negatives. Of the 3,149 benchmark-E2 pairs that the calibrated score moves to E0, 12.5\% have nonzero penetration, as do 8.6\% of the 535 E1 pairs moved there. In all, 3.9\% of the 11,550 pairs the calibrated score rates E0, about 450, show Claude usage. At the task level, 304 of the 1,288 calibration tasks with positive penetration (23.6\%) are rated E0 by $A_2$ and 159 (12.3\%) by $A_1$, and their penetration is not marginal: its median is 0.91, the same as for all positive tasks. Only 46 of the 304 are unanimous, with all seven models at E0. The rest are tasks on which the models split and the conservative weighting of the calibrated arm settles near zero. The examples with the highest penetration are tasks with a documentation or communication component inside work whose bottleneck is elsewhere: documenting patients' medical histories and assessment findings (critical care nurses, benchmark E1, $A_2$ 0.20, $A_1$ 0.36), recommending products to customers based on their needs (wholesale sales representatives, E2, 0.16, 0.36), preparing and delivering lectures on chemistry (postsecondary teachers, E2, 0.00, 0.29), and interviewing people and keeping track of their responses (statistical assistants, E2, 0.00, 0.00, with all seven models at E0). Observed use does not require that the task's time be halved, so these cases bound how literally a zero should be read rather than showing the rubric to be wrong, but they also show that the calibrated arm's zeros are the least reliable part of the index.

Table~\ref{tab:label_examples} lists examples of the three large flows, spread across \onet major occupational groups. The pattern is consistent. Tasks moved from E1 to E0 are dominated by ``report'', ``announce'', ``explain'', and ``provide'' verbs, whose time cost comes from being physically present in real time. Tasks moved from E2 to E0 are ones for which the benchmark imagined tooling that would close the gap, but for which the evidence records no deployed system doing site inspection, live negotiation, or in-person presentation. Tasks raised from E2 to E1 are document, research, and analysis work that a browser-and-workspace agent can do today without further integration. The smallest flow, the 13 pairs raised from benchmark E0 to grounded E1, contains several apparent errors. One example is maintaining a license, whose binding cost is the process of the issuing authority rather than the paperwork. At 0.07\% of matched pairs, this flow does not affect any aggregate, but it clearly shows where the measure still misreads a task.

\begin{table}[htbp]
\centering
\small
\setlength{\tabcolsep}{4pt}
\begin{tabular}{>{\raggedright\arraybackslash}p{4.0cm}>{\raggedright\arraybackslash}p{8.0cm}cc}
\toprule
Occupation & Task & $\beta^\star$ & Claude use \\
\midrule
\multicolumn{4}{@{}l}{\emph{Benchmark E1 (directly exposed) $\rightarrow$ grounded E0, $n=535$}} \\
\midrule
Firefighters & Assess fires and situations and report conditions to superiors to receive instructions, using two-way radios. & 0.00 & 0 \\
Phlebotomists & Explain fluid or tissue collection procedures to patients. & 0.00 & 0 \\
Flight Attendants & Announce flight delays and descent preparations. & 0.00 & 0 \\
Special Education Teachers, Elementary & Establish and communicate clear objectives for all lessons, units, and projects to students. & 0.00 & 0 \\
Telephone Operators & Observe signal lights on switchboards, and dial or press buttons to make connections. & 0.00 & 0 \\
\midrule
\multicolumn{4}{@{}l}{\emph{Benchmark E2 (tool-dependent) $\rightarrow$ grounded E0, $n=3{,}149$}} \\
\midrule
Architects, Except Landscape and Naval & Conduct periodic on-site observations of construction work to monitor compliance with plans. & 0.00 & 0 \\
Lawyers & Present and summarize cases to judges and juries. & 0.00 & 0 \\
Compensation and Benefits Managers & Negotiate bargaining agreements. & 0.00 & 0 \\
Chiropractors & Advise patients about recommended courses of treatment. & 0.00 & 0 \\
Construction Managers & Inspect or review projects to monitor compliance with building and safety codes or other regulations. & 0.00 & 0 \\
\midrule
\multicolumn{4}{@{}l}{\emph{Benchmark E2 (tool-dependent) $\rightarrow$ grounded E1, $n=1{,}375$}} \\
\midrule
Market Research Analysts and Marketing Specialists & Gather data on competitors and analyze their prices, sales, and method of marketing and distribution. & 1.00 & $>0$ \\
Judges, Magistrate Judges, and Magistrates & Read documents on pleadings and motions to ascertain facts and issues. & 1.00 & $>0$ \\
Lawyers & Prepare legal briefs and opinions, and file appeals in state and federal courts of appeal. & 1.00 & $>0$ \\
Environmental Economists & Develop economic models, forecasts, or scenarios to predict future economic and environmental outcomes. & 0.96 & $>0$ \\
Web Administrators & Track, compile, and analyze Web site usage data. & 0.92 & $>0$ \\
\bottomrule
\end{tabular}
\caption{Examples of label flows between the benchmark and the calibrated grounded measure. $\beta^\star$ is the calibrated score and the final column reports whether the occupation--task pair has nonzero Anthropic task penetration.}
\label{tab:label_examples}
\end{table}

\FloatBarrier
\section{Pairwise Adjudication Exercises}
\label{app:adjudication}

During development we asked three frontier LLM judges outside the ensemble, DeepSeek-V4-Pro \cite{deepseek2026v4}, Nemotron-3-Ultra-550B \cite{nvidia2026nemotron3ultra}, and GLM-5.2 \cite{zai2026glm52}, and four human annotators, three of them not involved in the project, to rule on tasks where our label and a benchmark label disagreed. Each adjudicator saw the occupation, the task, the 2026 rubric, and the two candidate labels in random order, without knowing which system produced which. We disclose these exercises because they were run, but we do not report their outcomes as evidence, for three reasons. The item sets were built from an earlier vintage of the index (June 2026) and were curated toward tasks expected to be easy and relatable for general annotators and toward cases in which our label had a visible advantage, so any preference measured on them is favorable to our labels by construction and says nothing about the population of disagreements. The benchmark side of each item was the rounded mean of the published occupation-level $\beta$ over the occupations that contain the task, not the GPT-4 task label. That quantity lies on the label grid for only 7.9\% of tasks and correlates at 0.63 with the task labels. And the judges are themselves 2026 models that share the ensemble's rubric, so their agreement is also consistent with our labels matching what a 2026 model believes under that rubric \cite{zheng2023judging}. A comparison that could support a claim about label quality would sample contested tasks at random from the final index, use the GPT-4 task labels as the benchmark side, swap the order of the candidates across passes, and be labeled by annotators with experience in the occupations concerned.

\FloatBarrier
\section{Example Reasoning Traces}
\label{app:outputs}

Table~\ref{tab:reasoning_traces} shows selected rationales from one ensemble model. They illustrate the subtask decomposition, the use of the descriptor panel, and how the retrieved sources are cited or declared irrelevant. Some subtask detail in the middle of each rationale is omitted for space.

\begin{table}[htbp]
\centering
\small
\setlength{\tabcolsep}{4pt}
\begin{tabular}{>{\raggedright\arraybackslash}p{2.1cm} >{\raggedright\arraybackslash}p{3.0cm} >{\raggedright\arraybackslash}p{8.3cm} c}
\toprule
Occupation & Task & Reasoning & Label \\
\midrule
Software Developers & Modify existing software to correct errors, adapt it to new hardware, or upgrade interfaces and improve performance. & The task involves modifying existing software to correct errors, adapt to new hardware, or improve performance. Subtasks include: (1) diagnosing issues via debugging and testing, (2) coding modifications and refactoring, (3) testing and debugging the updated software, [...]. However, diagnosing complex errors and adapting to hardware may require deeper integration or human expertise. Overall, the dominant time bottleneck is coding and refactoring, which E1 can substantially accelerate with LLMs for code generation and bug fixes. Thus, the task is E1. & E1 \\
\addlinespace
Sales Managers & Monitor customer preferences to determine focus of sales efforts. & The task involves gathering and analyzing customer data to inform sales strategies. Subtasks include: 1) Collecting data via surveys or market research (E1: AI can automate data collection and initial analysis). [...]. The dominant bottleneck is data integration from internal systems, which may not be fully automated with general AI due to proprietary constraints. Thus, overall exposure is E2. O*NET indicates frequent written communication and moderate automation, supporting the need for AI assistance but not full automation. No relevant NEWS context provided. & E2 \\
\addlinespace
Neuro\-psychologists & Provide education or counseling to individuals and families. & The task of providing education or counseling to individuals and families by a neuropsychologist involves several subtasks: (1) conducting initial assessments to diagnose neuropsychological conditions, (2) developing and implementing treatment plans, (3) educating clients on [...], requiring in-person presence and real-time interaction that the agentic system cannot replicate. Thus, the overall task is dominated by E0 subtasks, leading to E0 exposure. NEWS context is not relevant as the sources discuss career counseling in general, not neuropsychology-specific tasks, so relevant\_context is false and supporting\_sources is empty. & E0 \\
\addlinespace
Pest Control Workers & Record work activities performed. & The task of documenting work hours or activities for pest control workers involves recording time, tasks performed, and potentially customer interactions. Breaking this down into subtasks: (1) logging into a time-tracking system, (2) entering detailed descriptions of work [...] but this specific documentation task lacks physical components. The automation level (26\% slightly automated) suggests manual processes, supporting high automation potential. No image capabilities are needed, as the task is text-based. Thus, the overall task is E1: direct exposure, as the agent can cut time by over 50\% using general capabilities. & E1 \\
\bottomrule
\end{tabular}
\caption{Selected occupation--task reasoning traces from OpenReasoning-Nemotron 32B with predicted exposure labels.}
\label{tab:reasoning_traces}
\end{table}

\end{document}